\documentclass[11pt,a4paper,twoside, openright]{book} % Default font 11pt, A4, all pages are printed the same, new chapter always begins on a right-hand page.

\usepackage{natbib}
\usepackage[english]{babel} 
\usepackage[applemac]{inputenc} % Macintosh users should include this package instead of ansinew or latin1 for correct handling of diacritics.
\usepackage[T1]{fontenc}
\usepackage{mathptmx} 
\usepackage{helvet}
\usepackage{dpfloat}

\usepackage{url}
\def\apjl{ApJ}%
\def\apjs{ApJS}%
\def\ao{Appl.~Opt.}%
\def\aap{A\&A}%
\def\mnras{MNRAS}%
\usepackage{ThesisSU} % This package is specific for theses written at Stockholm university. Modifications to the classfile and the document can be found here.
\ifpdf
   \usepackage[pdftex]{graphicx}
   \usepackage{ifpdf}
   \usepackage[final]{pdfpages}
  \else
    \usepackage[dvips]{graphicx}
\fi % Used for figures

\usepackage[pdftex]{graphicx}
\usepackage{color}
\usepackage{amssymb}
\usepackage{fancyheadings}                      

\usepackage[colorlinks=true, urlcolor=blue, pagecolor=black, linkcolor=black, citecolor=black, filecolor=black, menucolor=black, pdfpagelayout=TwoColumnRight, pdfstartview=FitH, plainpages=false, pdfpagelabels]{hyperref}                   
                         
\def\sun{\hbox{$\odot$}}

\DeclareMathAlphabet{\mathsc}{OT1}{cmr}{m}{sc}
\def\testbx{bx}%
\DeclareRobustCommand{\ion}[2]{%
\relax\ifmmode
\ifx\testbx\f@series
{\mathbf{#1\,\mathsc{#2}}}\else
{\mathrm{#1\,\mathsc{#2}}}\fi
\else\textup{#1\,{\mdseries\textsc{#2}}}%
\fi}

\newcommand{\topTextDoc}{Doctoral Dissertation 2010}
\newcommand{\myInst}{Department of Astronomy}
\newcommand{\myInstB}{Stockholm University}
\newcommand{\instAdress}{SE-106 91 Stockholm}

\newcommand{\authorSurname}{de la Cruz Rodr\'iguez} % Your surname
\newcommand{\authorFirstName}{Jaime} % Your given name
\newcommand{\dissertationTitle}{Measuring the solar atmosphere} % The title of the dissertation
\newcommand{\dissertationSubtitle}{}% The subtitle of the dissertation (if there is any).
\newcommand{\yearOfPublication}{2010} % Year of publication
\newcommand{\placeOfPublication}{Stockholm} % The place of publication
\newcommand{\ISBN}{978-91-7447-154-0} % The ISBN number of the dissertation.
\newcommand{\distributor}{Department of Astronomy, Stockholm University} % Distributor

\newcommand{\printer}{Universitetsservice, US-AB}  % Printer
\newcommand{\printingdate}{Stockholm 2010} % Year and place of printing

\newcommand\abstracttext{%
\noindent
\section*{Abstract}
The new CRISP filter at the Swedish 1-m Solar Telescope provides opportunities for observing the solar atmosphere with unprecedented spatial resolution and cadence. In order to benefit from the high quality of observational data from this instrument, we have developed methods for calibrating and restoring polarized Stokes images, obtained at optical and near infrared wavelengths, taking into account field-of-view variations of the filter properties.

In order to facilitate velocity measurements, a time series from a 3D hydrodynamical granulation simulation is used to compute quiet Sun spectral line profiles at different heliocentric angles. The synthetic line profiles, with their convective blueshifts, can be used as absolute references for line-of-sight velocities.

Observations of the \ion{Ca}{ii} 8542 \AA \ line are used to study magnetic fields in chromospheric fibrils. The line wings show the granulation pattern at mid-photospheric heights whereas the overlying chromosphere is seen in the core of the line. Using full Stokes data, we have attempted to observationally verify the alignment of chromospheric fibrils with the magnetic field. Our results suggest that in most cases fibrils are aligned along the magnetic field direction, but we also find examples where this is not the case.

Detailed interpretation of Stokes data from spectral lines formed in the chromospheric data can be made using non-LTE inversion codes. For the first time, we use a realistic 3D MHD chromospheric simulation of the quiet Sun to assess how well NLTE inversions recover physical quantities from spectropolarimetric observations of \ion{Ca}{ii} 8542 \AA. We demonstrate that inversions provide realistic estimates of depth-averaged quantities in the chromosphere, although high spectral resolution and high sensitivity are needed to measure quiet Sun chromospheric magnetic fields.

} % The abstract text comes here. Not more than 300 words.

\newcommand\abstracttextSpikblad{%
\noindent
\textbf{Abstract:}
} % The 'spikblad' abstract text comes here. (not more than 300 words?)

\renewcommand{\listofpapers}
{\chapter*{List of Papers}
\noindent{\timesTen This thesis is based on the following publications:

} \vspace{13pt}
	
    \begin{romanlist}
\item \textbf{Solar velocity references from 3D HD photospheric models}

de la Cruz Rodr\'iguez J., Kiselman D., Carlsson M., 2010, submitted to A\&A
\\
\vspace{3mm}
\item \textbf{Non-LTE inversions from a 3D MHD chromospheric model}

de la Cruz Rodr\'iguez J., Socas-Navarro H., Carlsson M., Leenaarts~J., 2010, to be submitted to A\&A
\\
\vspace{3mm}
\item \textbf{Are solar chromospheric fibrils tracing the magnetic field?}

de la Cruz Rodr\'iguez J., Socas-Navarro H., 2010, submitted to A\&A
\\
\vspace{3mm}
\item \textbf{Stokes imaging polarimetry using image restoration at the Swedish 1-m Solar Telescope II: A calibration strategy for Fabry-P\'{e}rot based instruments}

Schnerr R., de la Cruz Rodr\'iguez J., van Noort M. 2010,  submitted to A\&A
\\
  \end{romanlist}
\vspace{13pt}
\noindent {\timesTen The articles are referred to in the text by their Roman numerals.}}

\newcommand{\dedication}%
{\newpage
\thispagestyle{empty}
\vspace*{\stretch{3}}
\begin{flushright}		
		{\fontfamily{pzc}\Large\selectfont
        \emph{A Klara}}

\end{flushright}
\vspace*{\stretch{1}}} % Dedication is optional. You can enter whatever feels right and use the fonts and images you like as long as they are embedded in the document. The suggested format does not have to be used.

\begin{document}

\thispagestyle{empty}
\frontmatterSU
\thispagestyle{empty}

\listofpapers
\thispagestyle{empty}

\tableofcontents

\thispagestyle{empty}
%\listoffigures
\thispagestyle{empty}
%\listoftables

\mainmatter
\chapter{Introduction}\label{intro}
The solar atmosphere constitutes a remarkably complex astrophysical laboratory that continuously performs experiments for us to observe. One reason for trying to understand this complicated environment is to establish with high precision simple but fundamental properties of the Sun. An important example is its chemical composition which can be put in context with our understanding of the astrophysical processes in the interior of the Sun and other stars, in the Galaxy, and in the early universe. To accomplish this, we need to take into account the dynamics of the solar photosphere as well as the physical processes involved in the formation of the spectral lines from which chemical abundance ratios are determined. This represents a major challenge and our confidence in the results rely heavily on the accuracy of measurements made with modern solar telescopes.

A second reason for studying the solar atmosphere, and one that is even more relevant in the present thesis, is to understand the mechanisms that generate the observed fine structure, dynamics, and magnetic field and to carry over that understanding to other astrophysical plasmas, including other stellar atmospheres. Dramatic progress in this field has in recent years been made in part by improved theoretical simulations and in part by new solar telescopes, equipped with powerful instrumentation, on the ground and in space. Both simulations and observations clearly demonstrate that much dynamics occur at very small spatial scales. Obtaining accurate quantitative information that will allow us to confirm or refute new models requires pushing existing telescopes to their diffraction limit and designing future telescopes with improved spatial resolution, better signal-to-noise and equipped with multiple instruments to simultaneously diagnose different layers of the solar atmosphere. In addition, sophisticated post-processing methods are needed to enhance the fidelity of the observed data, by developing accurate methods for calibrations and for removing contamination from straylight due to limitations set by the Earth's atmosphere and optical imperfections in the telescope or its instrumentation. This thesis deals with the challenges of accurate measurements of quantities relevant to small-scale dynamics, based on observations with a major solar telescope: the Swedish 1-m Solar Telescope (SST) on La Palma.

Our observational data are from two distinct, but physically connected, layers of the solar atmosphere: the photosphere and the chromosphere. The dynamics and morphology of these two atmospheric layers are completely different. To a large extent, these differences can be attributed to magnetic fields: whereas the gas pressure falls of exponentially with height and is roughly $10^5$ times smaller in the chromosphere than in the photosphere, the magnetic field strength falls off much slower. The relative importance of forces associated with gas pressure and  magnetic field can be estimated from the ratio of gas pressure ($P_g$) to magnetic pressure ($P_B$), the plasma-beta parameter defined by:
$$
\beta = \frac{P_g}{P_B}
$$
In the photosphere, $\beta$ is much larger than unity everywhere, except in sunspots and other (mostly small-scale) concentrations of magnetic field.  The photosphere is dominated by a convective energy flux, peaking just below the visible surface. Key questions today are to understand how this energy flux is maintained within magnetic structures and a major challenge is to identify and measure the velocity signatures of any convective flows present. These signatures are both weak and and small-scale and their identification relies on whether we can establish an absolute reference for measured Doppler velocities on the Sun. The first part of the present thesis deals with this problem.

The second part deals with observations of the chromosphere. Here, magnetic fields are much weaker than in the strongest magnetic structures seen in the photosphere but the gas pressure is even lower. In the upper chromosphere, $\beta<1$ thus the magnetic forces are dominant.  This work aims at investigating the potential for diagnosing the weak chromospheric magnetic fields using Stokes polarimetry and sophisticated inversion techniques.\\

To set the context, we proceed with an overview of some photospheric and chromospheric topics.

\section{The Photosphere}\label{photos}
The visible surface of the Sun corresponds to the photosphere, a thin layer of about 500 km located on top of the convection zone, where the plasma changes from completely opaque to transparent \citep{2002stix}. 
\subsection{Granulation}\label{igran}
\begin{figure}[]
      \centering
    %  \resizebox{\hsize}{!}{\includegraphics[trim=0 0.1cm 0.2cm 0.1cm, clip]{figures/chap1_quiet.pdf}}
         \resizebox{\hsize}{!}{\includegraphics[trim=0.15cm 0.4cm 0.5cm 0.3cm, clip]{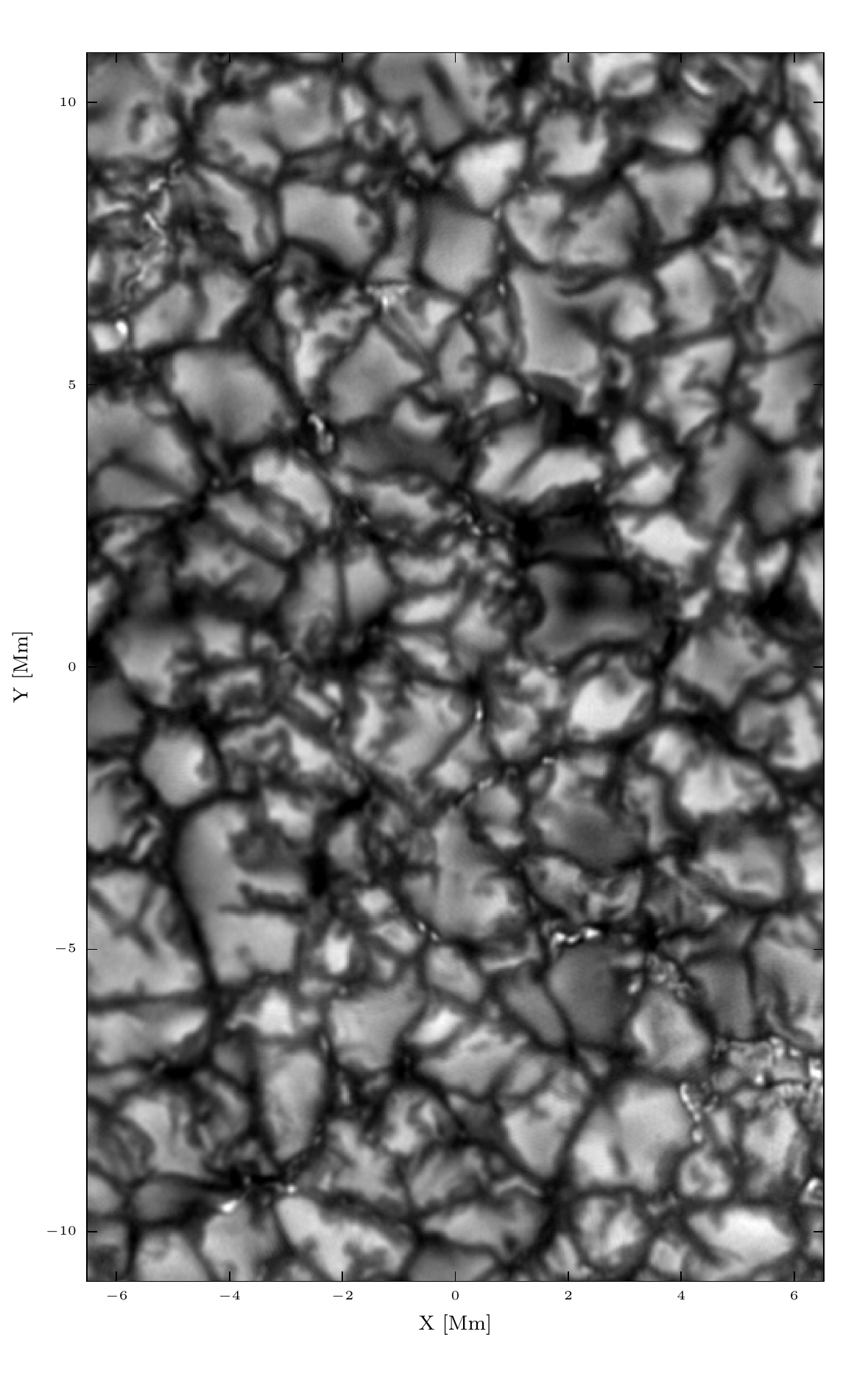}}

        \caption{
		The photosphere imaged in the G-band at 430 nm with the Swedish 1-m Solar telescope. Image courtesy of M. van Noort \& L. Rouppe van der Voort (ITA-UiO).
       }
        \label{qtra}
\end{figure}

Outside active regions with strong magnetic fields, the  photosphere is dominated by a dynamic pattern of bright \emph{granules} surrounded by dark \emph{intergranular lanes} (see Fig. \ref{qtra}). The flow in a granule resembles that of a fountain, where the hot plasma moves upwards inside the granules and then flows out towards the edge, where the cooler plasma merges with material from neighbouring granules. Gravity and pressure increase at the edge of the granules, accelerating the fluid downwards \citep{1998stein}. Regular granules have a typical size of the order of 1 Mm, and a characteristic life time of 6  minutes \citep{1961bahng}. 

Convective motions leave strong fingerprints on any line formed in the photosphere. An important diagnostic is the C-shaped bisector obtained from spatially-averaged line profiles. This effect is produced by the statistical average of bright blueshifted profiles from granules with dark redshifted profiles originating in the intergranular lanes \citep{1981dravins}. This intensity weighted average is blueshifted as upflows are more heavily weighted by being brighter and covering a larger area than the narrower intergranular lanes. This shift is commonly known as the  \emph{convective blueshift}.

\begin{figure}[]
      \centering
      \resizebox{\hsize}{!}{\includegraphics[]{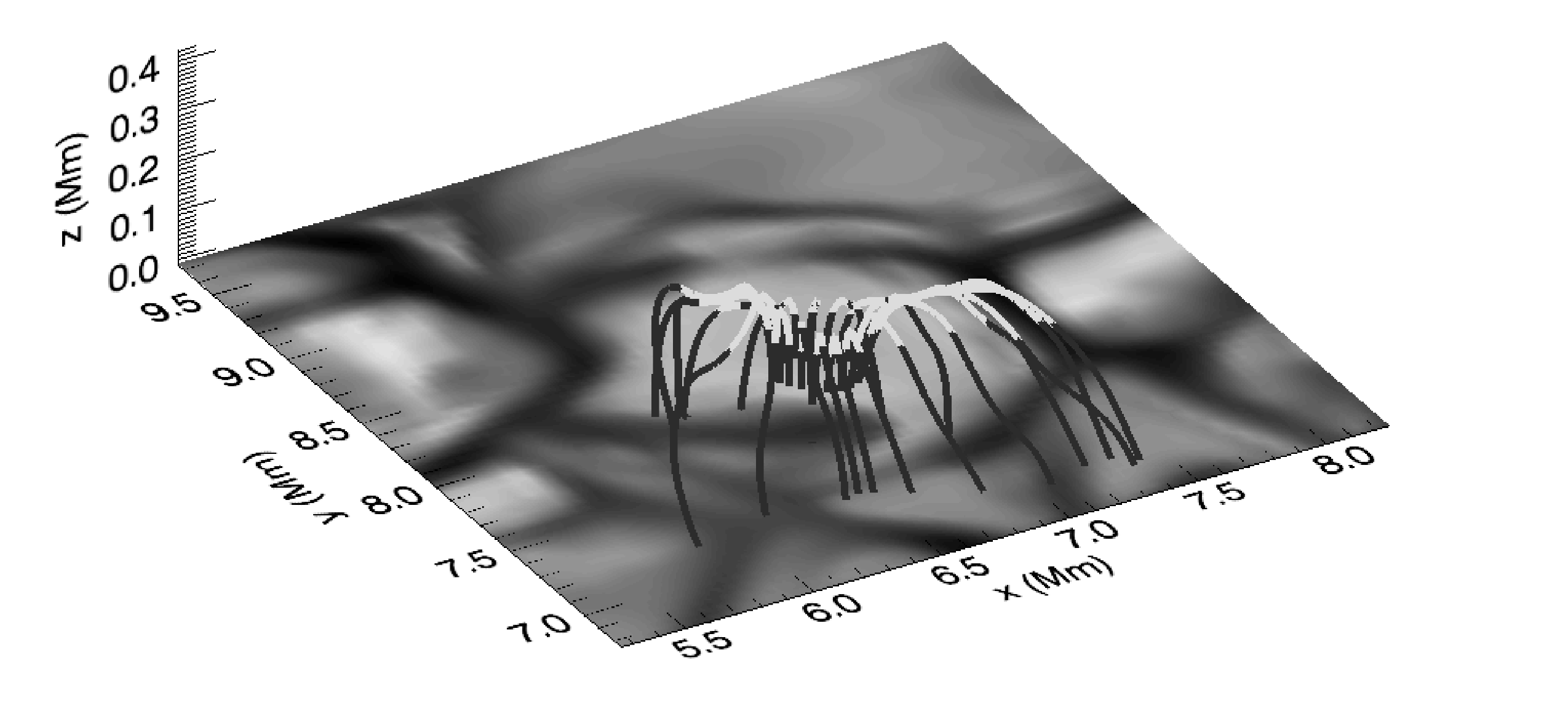}}
        \caption{
		Trajectories of tracer fluid elements above a granule. The grayscale image show the vertical velocities at at $Z=0$ when the tracers were released. The colors in the trajectories correspond to $Q_{\mbox{rad}}<0$ (dark-grey) and $Q_{\mbox{rad}}>0$ (light-grey). From \citet{2007cheung} (reproduced with permission of the authors).
       }
        \label{fcheung}
\end{figure}

The thermodynamic history of fluid elements rising through the photosphere is described in detail by \citet{2007cheung}. The temperature decrease of a fluid element that moves up  in the convection zone, is mostly produced by adiabatic expansion. As the fluid reaches the photosphere, the opacity drops and the fluid rapidly loses entropy by radiative cooling. At this point, the fluid cell is overshooting into the stably stratified photosphere, and it still interacts with the surroundings because it is not completely transparent to radiation. Fig. \ref{fcheung} shows the trajectory described by tracer fluid elements that enter the photosphere. The color coding indicates the sign of $Q_{\mbox{rad}}$, the heat exchange with the surroundings, in dark for radiative loses ($Q_{\mbox{rad}} <0$) and in light grey where the fluid elements are being heated ($Q_{\mbox{rad}}>0$). Interestingly, there is no direct correlation between heat exchange and the temperature variation of the fluid elements, which is determined by a balance between adiabatic expansion and (non-adiabatic) heat exchange with the surroundings. 

In recent years, efforts to obtain refined estimates of solar abundances have given rise to some controversies \citep[see e.g.][]{2000asplund2}.  This debate stimulated improvements of treatment of energy transfer by radiation in 3D hydrodynamic simulations, in particular for the mid and high photosphere, where spectral lines are formed. These controversies have also stimulated the development of improved non-LTE calculations of spectral lines used for abundance calculations \citep{2001trujillo}. As a result, 3D MHD simulations of more complex solar atmosphere dynamics involving magnetic fields can now also be made with improved energy transfer than just a few years ago.

\subsection{Sunspots}
The structure and dynamics of sunspots remain some of the most controversial topics in the solar photosphere. Sunspots constitute strong magnetic field concentrations that appear in the atmosphere and present typical sizes of 12000 km. As a first approximation, we can assume that the magnetic field acts on the gas in the form of a magnetic pressure, 
$$
P_B = \frac{B^2}{2\mu_0}.
$$
Horizontal force balance then dictates that the sum of the gas pressure and magnetic pressure must be the same inside the sunspot as outside. This implies that
$$
 P_s + \frac{B^2}{2\mu_0} = P_{qs}
$$
where $P_s$ is the gas pressure inside the spot and $P_{qs}$ is the gas pressure in the surrounding (non-magnetic) quiet sun. An immediate consequence of this is that the gas pressure inside the spot must be lower than outside and that therefore the atmosphere is more transparent inside sunspots,  allowing observers to see deeper layers of the atmosphere than in quiet Sun observations. This is generally known as the Wilson depression ($\sim500$ km), discovered by \citet{1774wilson}. At the same time the $H^-$ opacity decreases with temperature, and sunspots are cooler than their environment so the opacity decreases even more. Sunspots are darker than their surrounding granulation because convection is suppressed by the strong magnetic field, thus sunspots are cooler as an effect of inefficient heat transfer.

\begin{figure}[!h]
      \centering
       \resizebox{\hsize}{!}{\includegraphics[trim=0.48cm 0.7cm 0.1cm 0.55cm, clip]{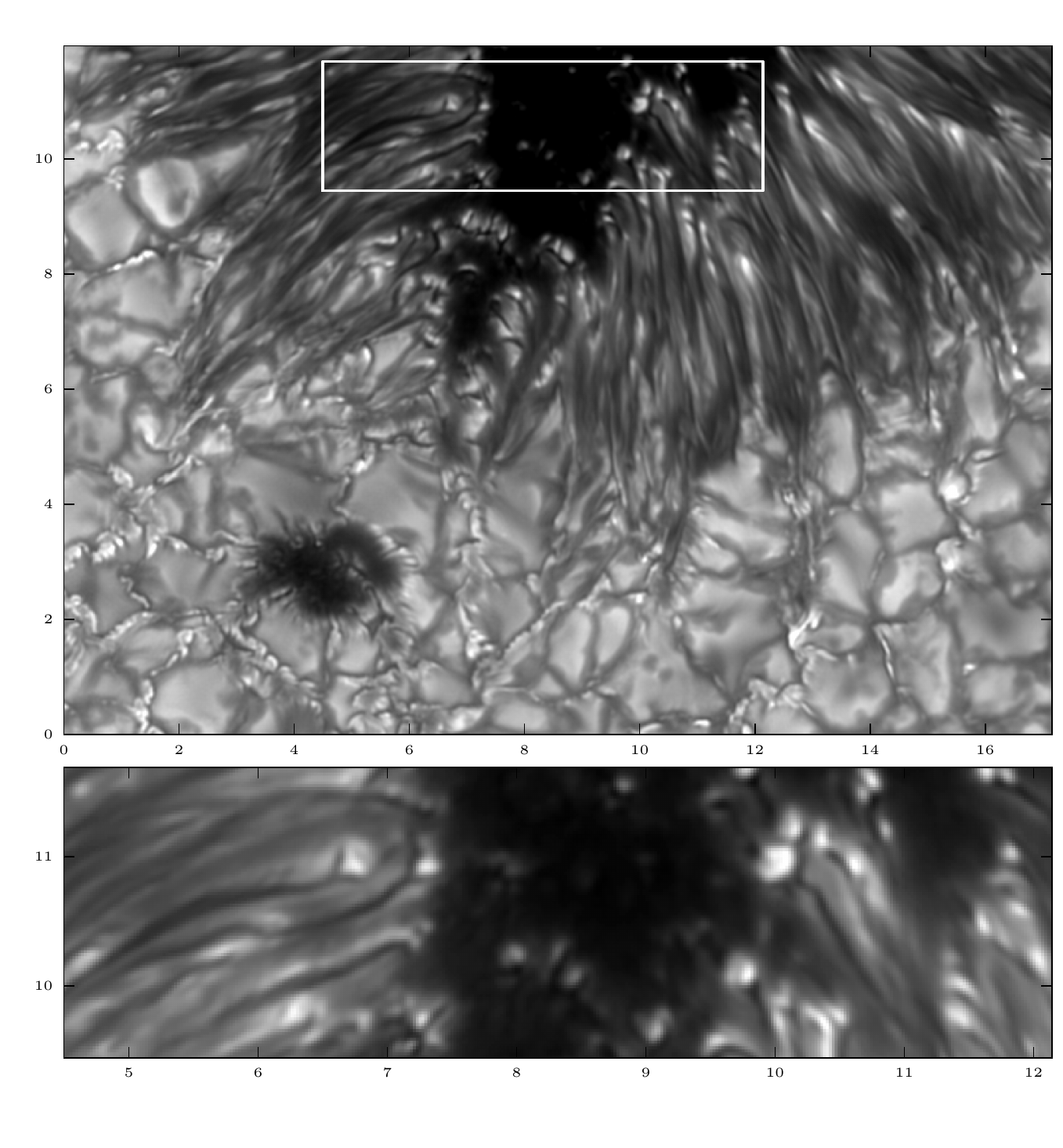}}
        \caption{Surroundings of a sunspot (top) and close-up view of penumbral filaments (bottom). The units of the axis are given in Mm. Continuum observations at 396 nm, by Vasco Henriques (ISP-KVA).
       }
        \label{sunsvas}
\end{figure}

During the past decade, the scientific debate has focused on the dynamics and structure of the penumbra and explaining why the penumbra is as bright as it is, about 75\% of the surrounding quiet sun. Recent advances in instrumentation have unveiled very fine structure in sunspots, especially in the penumbra. Fig. \ref{sunsvas} illustrates in great detail the fine structure of the penumbra of a sunspot. The blown-up section shows dark cored penumbral filaments \citep{2002scharmer} and the inner umbra. 
The following theoretical frameworks \citep{2008scharmerR} have become popular because they can partially reproduce the features observed in sunspots, in spite of representing  different physical concepts. 

\begin{enumerate}
	\item \textbf{The uncombed penumbra model} \citep{1993solanki} postulates the existence of discrete flux tubes with homogeneous magnetic field inside that discontinuously changes at the boundary. Those nearly horizontal flux tubes are embedded in a more vertical magnetic field. This model was able to reproduce strongly asymmetric Stokes V profiles observed on the limb side of the penumbra in observations off disk center \citep{1992sanchez-almeida}.\label{uncomb}
	\item \textbf{Siphon flow models} \citep{1968meyer,1997montesinos} are based on the idea that a difference in field strength between the two footpoints of a flux tube leads to a difference in gas pressure, driving a plasma flow in the direction of the footpoint with the highest field strength (thus lower gas pressure). Evershed flows are assumed to be steady flows between two footpoints with different magnetic field strengths \citep[e.g.,][]{1997westendorp-plaza}. However, these models do not explain the mechanism involved in such field strength difference at the footpoints.\label{siphon}
	\item \textbf{Convection and downward pumping of magnetic flux} are ingredients added to the siphon models. As siphon models present a stationary solution, time variations are explained by external mechanisms to the penumbra. In this context, moving penumbral grains are assumed to be produced by a moving convective pattern in the bright side of the penumbra.  \citet{2002thomas} and \citet{2004weiss} attribute the submergence of the flux tubes at the boundary of the penumbra to downward pumping produced by convective motions. Furthermore, they attribute the whole filamentary structure of the penumbra to the same downward pumping mechanism, explaining the structure inside the sunspot based on mechanisms that take place outside.\label{sipin}
	\item \textbf{The convective gap model} proposed by \citet{2006scharmerspruit}. In this framework, the proposed mechanism that generates penumbral filaments is convection, in radially aligned, nearly field free gaps. The strong field gradients that are necessary to reproduce the asymmetric Stokes V profiles reported by \citet{1992sanchez-almeida} are assumed to be produced by the topology of nearly field-free gaps combined with line-of-sight gradients in the flow velocity. The Evershed flow is explained as representing the horizontal component of this convection. \label{gappy}
\end{enumerate}
%\subsection{Photospheric abundances (remove?)}
%Numerical simulations have been used during the past decades as tools to understand physical problems on the sun. \citet{2000asplund2} approached the solar abundances problem with an innovative idea. Spatially resolved spectra from a given numerical simulation cannot be compared one-to-one with real observations, however statistically the agreement is remarkable. They calculated spatially averaged line profiles computed assuming Local Thermodynamic Equilibrium (LTE) from their numerical simulation with and compared them with observations. They used the abundance as a free parameter in order to archive the best fit between observations and their simulations. This approach works because the Doppler width of the the average spectrum and its asymmetry is produced by the statistical average of bright-blueshifted spectra from granules and dark-redshifted spectra from intergranular lanes. Refinements to the values of the Fe abundance were done in non-LTE by \citet{2001trujillo}.
%
%Nevertheless, the effort to improve abundances has lead to mayor improvements on 3D simulations that nowadays include remarkably realistic physics \citep[see i.e.,][]{2009pereira}.  
\subsection{Oscillations in the solar atmosphere}
Solar oscillations were discovered by \citet{1962dop} with a simple observational technique:  two simultaneous images were recorded in the blue and the red wings of a spectral line, respectively, and then subtracted. The resulting image contained intensity variations produced by the Doppler shift of the line. \citet{1961kahn} proposed that oscillations are sound waves trapped in the solar atmosphere. Towards the solar interior, the temperature and speed of sound increase with the variation of the refractive index, caused by the increased density. The wave is refracted until it starts to propagate upwards. The same process occurs above the photosphere where the waves are refracted back into the inner atmosphere. Observations contain an overlap of hundreds of modes of oscillation that effectively reach different depths. The oscillations in the photosphere typically have a 5-minute period and an amplitude around 1 $\mbox{km}\ \mbox{s}^{-1}$ \citep{2002stix}.

\section{The chromosphere}
The chromosphere represents many challenges for solar physicists. Despite important discoveries during the past decade, it still remains unexplored to a large extent. From an observational point of view, only a few spectral lines are sensitive to the chromospheric  height range and those are usually hard to model and understand, like \ion{H}{i} 6563 \AA, \ion{Ca}{ii} K \& H (3934 and 3968~\AA \ respectively),  the \ion{Ca}{ii} infrared triplet (8498, 8542 \& 8662~\AA), \ion{Na}{i} $\mbox{D}_1$ (5896~\AA) , \ion{He}{i} 10830~\AA.

During the past 15 years combined efforts from observational and computational approaches have lead to a better understanding of chromospheric dynamics. 3D simulations of solar-like stars including a chromosphere and corona are now computationally affordable \citep{2007leenaarts,2007hansteen,2010carlsson}. 

On the observational side, a new generation of Fabry-Perot interferometers (FPI), for example  IBIS at the Dunn Solar Telescope (DST) and CRISP at the Swedish 1-m Solar Telescope (SST), have provided evidences of very fine structure in the chromosphere.

\subsection{The chromospheric landscape}\label{chr_feat}
\begin{figure}[]
      \centering
      \resizebox{1.0\hsize}{!}{\includegraphics[trim=0.2cm 0.2cm 0.27cm 0.2cm, clip, angle = -90]{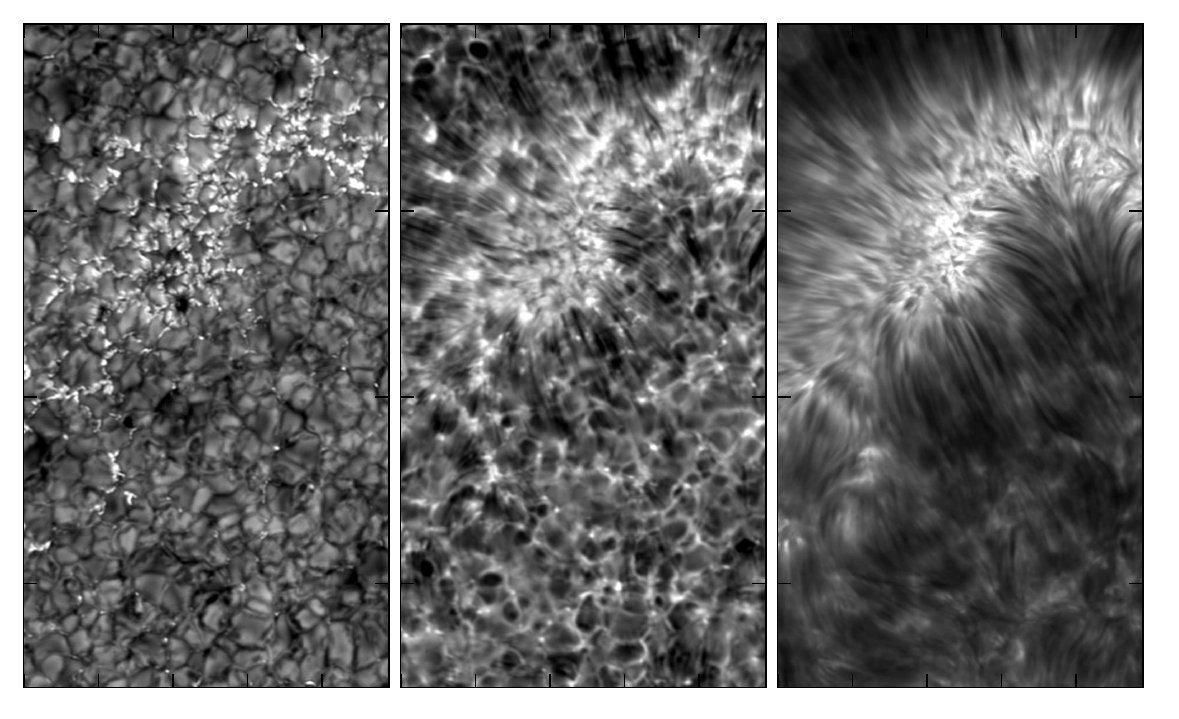}}
        \caption{
		Images in \ion{Ca}{ii} 8542 \AA \ taken at the SST. \emph{Top:} the mid-photosphere. \emph{Middle: } High-photosphere/low-chromosphere. \emph{Bottom:} the chromosphere. 
       }
        \label{ca2}
\end{figure}

The definition of chromospheric fine structure has evolved as new discoveries were made. It is widely accepted that the chromosphere includes the characteristic grass-like topology usually seen in $\mbox{H}_\alpha$ images \citep{2006rutten}, but it is not clear where the boundaries of the chromosphere are. Fig. \ref{ca2} shows three \ion{Ca}{ii} 8542 \AA \ images acquired with SST/CRISP. This strong spectral line shows photospheric granulation at the wings and chromospheric fibrilar features in the core. Some chromospheric features are: 
	\begin{figure}[]
      \centering
      \resizebox{\hsize}{!}{\includegraphics[]{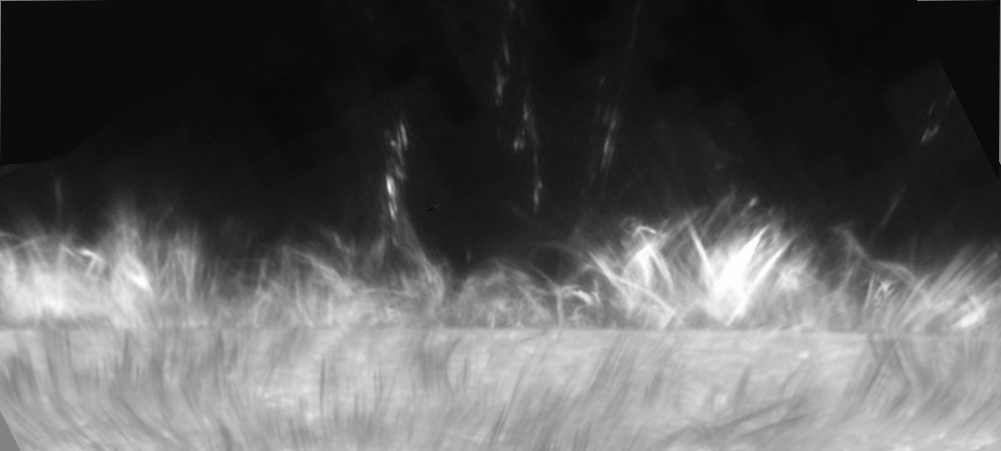}}
        \caption{
		Spicules at the limb seen at \ion{H}{i} 6563 \AA.  Intensity gradients towards the limb have been filtered to enhance the off-limb part. Image taken at the Swedish 1-m Solar Telescope, courtesy of  Luc Rouppe van der Voort (ITA-UiO).
       }
        \label{spic}
\end{figure}
\begin{itemize}
	\item \textbf{Straws} are bright features seen in the core of chromospheric lines \citep{2007rutten}. They start in facular regions in the photosphere and are much brighter than their surroundings in the chromosphere, showing hedge shapes in filtergrams. Fig. \ref{ca2} shows a close view of straws, . In the upper panel, photospheric faculae become brighter than the surroundings at mid-photospheric heights. Fibrils seem to originate in these straws, as shown in the lower panel in Fig. \ref{ca2}.
	\item \textbf{Fibrils,} also known as mottles, are elongated dark features that in \ion{H}{i}~6563~\AA  \ (hereafter $H_\alpha$) form a grass-like canopy covering internetwork cells at any heliocentric angle. They are also present \ion{Ca}{ii} images, but they only appear around network patches, as shown in Fig. \ref{ca2}. They are very dynamic and show transversal motions over a time period of  2 seconds. \citet{2006hansteen} and \citet{2007rouppe} proposed that dynamic fibrils are  driven by magneto-acoustic shocks that leak into the chromosphere along magnetic field lines.
	\item \textbf{Spicules:} limb images of the chromosphere are dominated by spicules. \citet{2007depontieu} provided a detailed description of spicules and proposed physical mechanisms that could drive them. Spicules appear as thin, long highly dynamic features, usually reaching heights around 5000~km (see Fig. \ref{spic}). Their width varies from 700~km down to current telescopes diffraction limit ($\sim 100$ km). Spicules are classified in type I and type II. Type I spicules move up and down in time scales of 3-7 minutes and some of them present transversal motions. However, those fibrils that do not move transversely show  acceleration and trajectories that are similar to those of dynamic fibrils, suggesting that they are also driven by magneto-acoustic shocks. Type II spicules are very dynamic and show apparent speeds between 50-150 $\mbox{km} \ \mbox{s}^{-1}$, disappearing in time scales of 5-20 s. The mechanism driving type II spicules is not well understood, although their rapid disappearance suggests that  strong heating could be ionizing \ion{Ca}{ii} atoms. Recently, \citet{2009rouppe} found on-disk counterparts of spicules, which produce a clear signature in the blue wing of the  \ion{H}{i} 6563 and \ion{Ca}{ii} 8542 lines.
	\item \textbf{Filaments} are (dark) cold clouds of material that according to their temperature, belong to the chromosphere. They present typical lengths of $200 000$~km, with thickness's of $5000$ km \citep{2002stix}. Towards the limb, filaments are known as \textbf{prominences} that appear \emph{hanging} above the chromosphere up to $50 \ 000$ km. The only known mechanism that can sustain such cold and dense material is an electromagnetic force. Photospheric observations show that filaments predominantly exist along neutral magnetic field lines. Present observational efforts aim at measuring magnetic field in the filaments. 
\end{itemize}

  \begin{figure}[]
      \centering
      \resizebox{\hsize}{!}{\includegraphics[trim=0 0 0.7cm 0]{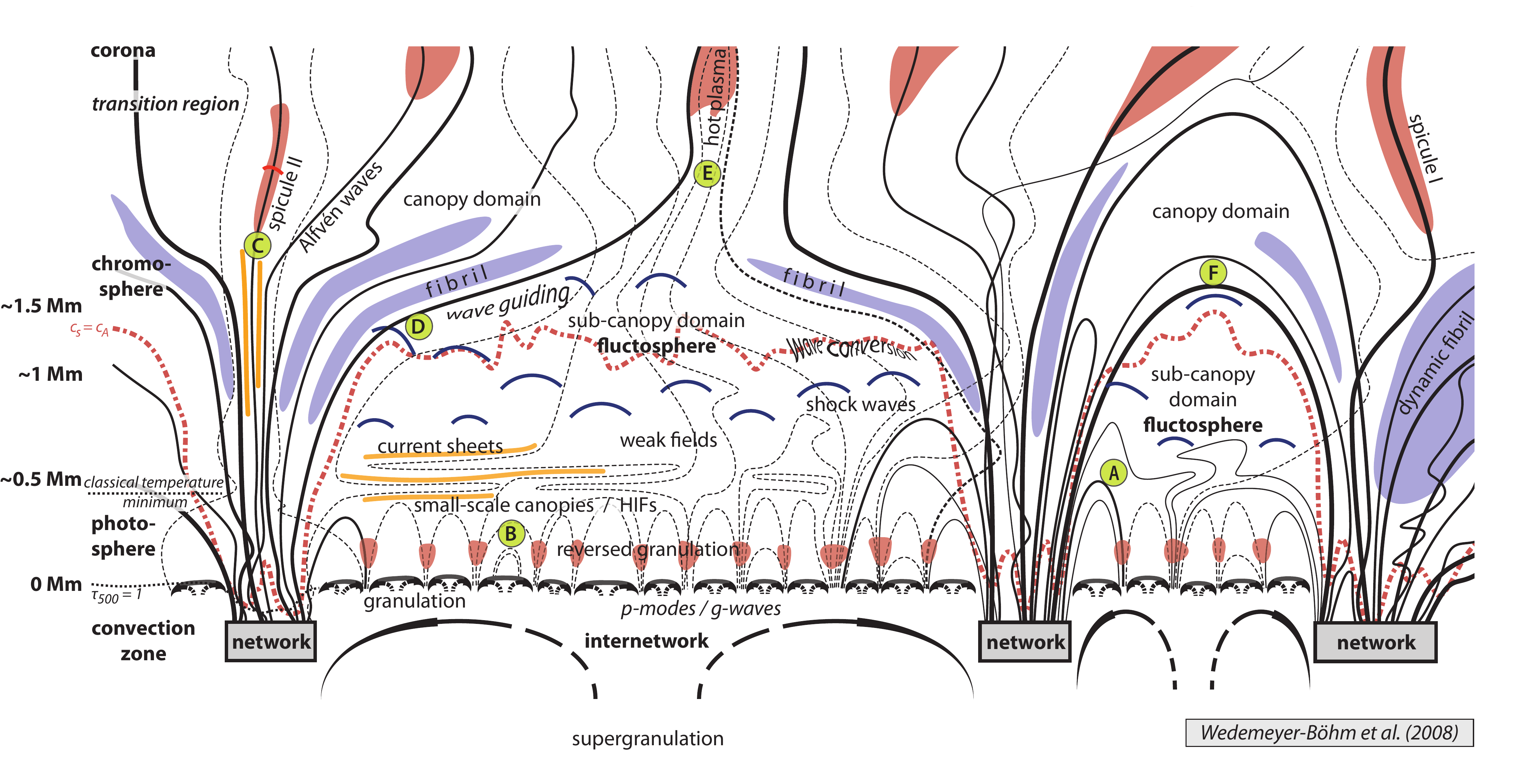}}
        \caption{
		Schematic structure of the quiet sun atmosphere \citep{2009wedemeyer}, including some network patches. The black solid lines represent magnetic field lines that are anchored in the photosphere through network regions.
       }
        \label{wedemeyer_sketch}
\end{figure}
\subsection{Chromospheric heating}\label{chr_heat}
 One outstanding question about the chromosphere relates to its energy budget. Why is the outer atmosphere of the sun hotter than the photospheric surface? Semi-empirical, one-dimensional models of quiet Sun require the average temperature to rise above the photosphere to reproduce the chromospheric intrinsic enhanced emission \citep[VAL3,][]{1981vernazza}. However, observations show that the chromosphere is  vigorously active and strongly inhomogeneous. \citet{1995carlsson} demonstrated that enhanced emission can be produced by shocks without increasing the mean gas temperature. At present, the connection of waves with chromospheric heating appears widely accepted \citep{2005fossum,2007cauzzi,2007wedemeyer,2009vecchio}, however the exact role of these waves is still under debate. 

 \subsection{Magnetic field configuration}\label{chr_field}
The relatively organized and elongated fibrils seen in chromospheric lines  (see  Fig. \ref{ca2}) suggests the presence of magnetic fields. However, if magnetic fields dominate the chromosphere with $\beta \ll 1$, those should be almost force free, leading to relatively smooth magnetic configuration. Thus, the complex fine structure must relate to the thermodynamics of the plasma \citep{2006judge}. 
 
The features described in \S\ref{chr_feat} and their connection with the photosphere have been contextualized by \citet{2009wedemeyer} in the cartoon shown in Fig. \ref{wedemeyer_sketch}. Magnetic field lines form a canopy where $\beta \sim 1$. Below the magnetic canopy, acoustic waves originating in the photosphere are dissipated, producing short-lived bright features seen in the wings of \ion{Ca}{ii} images. In this cartoon, fibrils and spicules form the magnetic canopy, which originates from network patches in the photosphere and extends over internetwork regions in the chromosphere.\\

\noindent We investigate the relation between fibrils and magnetic fields in \textbf{Paper III}, looking for an observational evidence of their alignment. 

\chapter{Velocity references on solar observations}\label{velref}
This chapter describes the  inherent difficulties in measuring absolute line-of-sight (LOS) velocities from spectroscopic observations. To illustrate the importance of this problem, we recall the discussion in Chapter \ref{intro} where we summarized the main theoretical frameworks that have been proposed to explain the structure and dynamics of sunspots. A key difference between fluxtube models and the field-free gap model is that the latter explains penumbral filaments as convective intrusions where magnetic field is weak enough not to suppress convection. Thus, observational evidence of convective motions inside penumbral filaments would be a key to explaining the origin of its filamentary structure and choosing among existing models. These types of measurements are very hard to make because in the upper part of the filament overshooting convection is expected to be weak. At the same time, the presence of the Evershed flow and the small scales involved, makes it very hard to establish the existence of overturning convection inside penumbral filaments. For these reasons, a very accurate velocity calibration is needed to make it possible to distinguish between the upflowing and downflowing components of any convection.

\section{The calibration problem}\label{prob}
The fundamental question that is addressed here is:
\emph{What defines the local frame of rest on the Sun?}
An observer placed on the Sun would apply Eq. \ref{dop} to measure LOS velocities, would apply the relationship
\begin{equation}
v = \frac{\lambda - \lambda_0}{\lambda_0}\cdot c,
\label{dop}
\end{equation}
where $\lambda$ is the observed wavelength, $\lambda_0$ is the reference wavelength (usually the laboratory wavelength of the line of interest) and $c$ is the speed of light.
However, ground-based observations are affected by the rotation of the Earth ($v_{\oplus,\textrm{rot}}$), the radial component of the Earth's orbital motion ($v_{\oplus,\textrm{orbit}}$), the rotation of the Sun ($v_{\sun,\textrm{rot}}$)  and the graviational redshift ($v_{\textrm{grav}}$), the relarion is in reality more complex,
\begin{equation}
v = \frac{\lambda - \lambda_0}{\lambda_0}\cdot c + v_{\oplus,\textrm{rot}} + v_{\oplus,\textrm{orbit}} + v_{\sun,\textrm{rot}} + v_{\textrm{grav}},
\label{dop_all}
\end{equation}
Furthermore the precision of the atomic data limits the accuracy of the conversion from wavelength to velocities, regardless of how accurate the instrument is. This calibration issue becomes more severe when observational practicalities make it difficult to compensate for the last three terms of Eq.~\ref{dop_all}. The gravitational redshift has a theoretical constant value of $633~\mbox{m} \ \mbox{s}^{-1}$ at the surface of the Earth \citep{2006cacciani}.

In addition, current instruments for solar observations seldom use laboratory light sources as reference wavelengths. The obvious solution of finding a $\lambda_0$ on the Sun itself is confounded by the convective lineshifts discussed in Sect.~\ref{igran}. The magnitude of these shifts is different for different lines.

Below, we summarize some of the methods that have been used to define a velocity reference for spectroscopic observations.\\

\begin{enumerate}
	\item\label{csuns}Convective motions in \textbf{sunspot umbrae} are suppressed by the presence of strong magnetic fields. Thus, it is common to assume that the umbra is at rest, defining a reference for line-of-sight velocities \citep[e.g.,][]{1977beckers,2008scharmer, 2010ortiz}. However, this assumption usually does not hold higher up in the chromosphere, where umbral flashes associated with shocks \citep{2000socas-navarro} produce strong blue-shifts. This approach carries the risk of being affected by spurious line shifts produced by molecular blends that only form in the umbra because it is colder than the surrounding granulation. Eq.~\ref{dop} can be used to compute the conversion from the wavelength scale to a velocity scale, but in this case $\lambda_0$ is the central wavelength of the spectral line measured in the umbra of a sunspot.
	\item \label{ctell}\textbf{Telluric lines} are sometimes present in the spectral range that is being observed. These lines are formed in the Earth's atmosphere. Thus, they allow the definition of a very accurate laboratory frame of rest that can be converted to the solar frame using ephemeris constants, the time of the observations and solar rotation (Eq. \ref{dop_all}). The conversion from wavelength to velocities is calculated using the laboratory wavelength of the line of interest. \citet{1997pillet} and \citet{2008bellot-rubio} used this approach to calibrate their observations.
	\item \label{catlas} \textbf{A spectral atlas} can be used to calibrate observations, as the effects of the rotation and translation of the Earth usually have been compensated for. \citet{2007langangen} used the atlas acquired with the Fourier Transform Spectrometer at the McMath-Pierce Telescope (hereafter FTS atlas) of \citet{fts-atlas} to calibrate some of his observations. This atlas was acquired at solar center, thus its usability is limited to disk center observations.
	\item \label{cmodel} \textbf{Numerical models} can be used to compute the convective shift of a line and use it as a reference for velocities. The advantage of this approach is that the convective shift is measured relative to the assumed laboratory wavelength of the line, so it is insensitive to uncertainties in the atomic data. \citet{2002bellot-rubio} computed a two-components model from the inversion of photospheric \ion{Fe}{i} lines. This model only contains the vertical component of the velocity field, thus it is limited to solar center. \citet{2004tritschler} and \citet{2009franz} used this model to calibrate disk center observations. However, \citet{2004bellot-rubio}, used it together with the empirical results of \citet{1988balthasar} to estimate lineshifts also off solar centre. \citet{2007langangen} used a 3D hydrodynamical simulation of solar convection to calibrate observations on \ion{C}{i} 5380 \AA. The central wavelength of this line is not known with enough precision to allow the use of the FTS atlas method that was used with their observations.
\end{enumerate}

\section{Calibration data from hydrodynamic granulation models}\label{method}
\begin{figure}[]
      \centering
       \resizebox{\hsize}{!}{\includegraphics[]{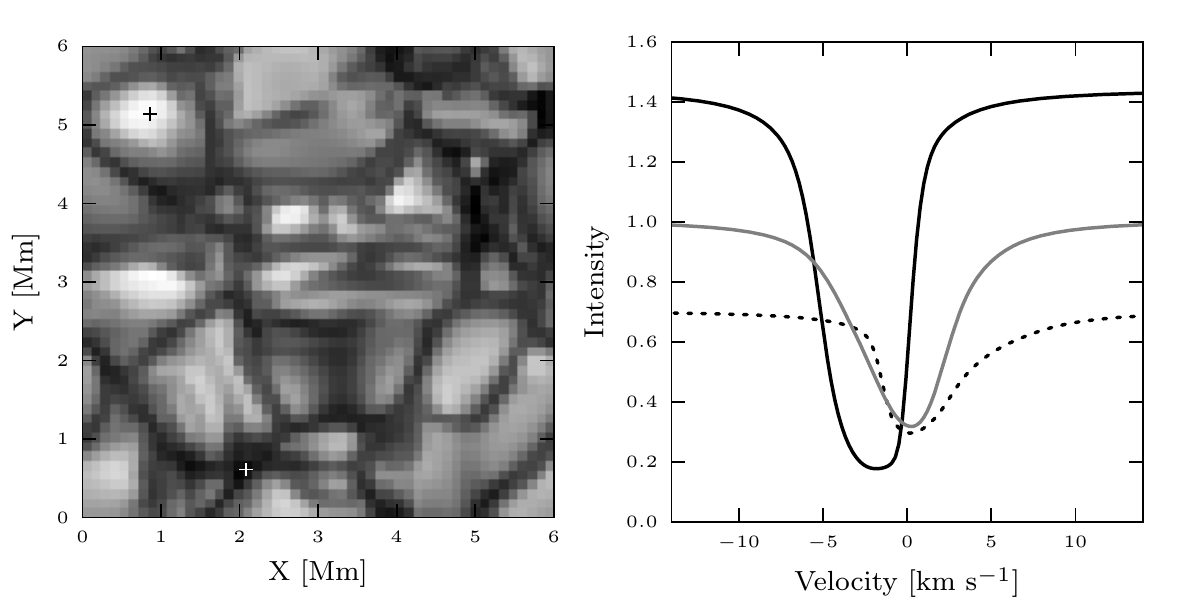}}
        \caption{\emph{Left:} Continuum intensity at 6301 \AA\ image from one of the 3D snapshots from the numerical simulation. The latter has been resampled at lower spatial resolution. \emph{Right:} Spectra from a blueshifted granule (solid black line) and redshifted intergranular lane (dashed line). The grey line corresponds to the spatial average of this snapshot. Two crosses on the left panel indicate the location of the spectra represented on the right panel. }\label{granmodel}
        \resizebox{0.9\hsize}{!}{\includegraphics[]{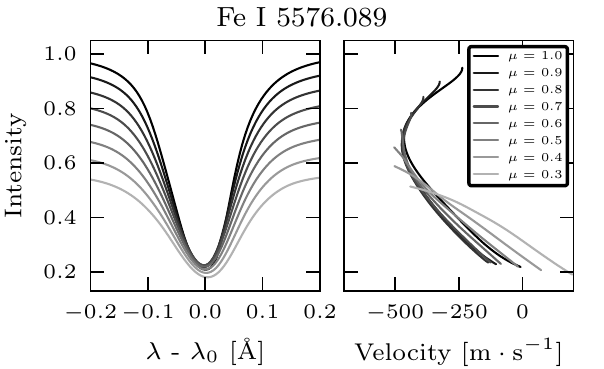}}
     %	\resizebox{0.89\hsize}{!}{\includegraphics[]{figures/c5380_vel.pdf}\includegraphics[]{figures/fe5250_vel.pdf}}	
     %	\resizebox{0.89\hsize}{!}{\includegraphics[]{figures/fe5251_vel.pdf}\includegraphics[]{figures/fe5576_vel.pdf}}
     %	\resizebox{0.89\hsize}{!}{\includegraphics[]{figures/fe6082_vel.pdf}\includegraphics[]{figures/fe6301_vel.pdf}}
     %	\resizebox{0.89\hsize}{!}{\includegraphics[]{figures/fe6302_vel.pdf}\includegraphics[]{figures/fe7090_vel.pdf}}
     %	 \resizebox{0.89\hsize}{!}{\includegraphics[]{figures/ca8498_vel.pdf}\includegraphics[]{figures/ca8542_vel.pdf}}
     %	 \resizebox{0.445\hsize}{!}{\includegraphics[]{figures/ca8662_vel.pdf}}
       % \caption{Synthetic profiles and the corresponding bisectors resulting from the calculations of \ion{C}{i}, \ion{Fe}{i} and \ion{Ca}{ii} lines. The grey scale indicates the variation of the heliocentric angle from $\mu=1.0$ to $\mu=0.3$. Note that the  bisectors for the \ion{Ca}{ii} lines are not shown in the core part.}
       \caption{Synthetic profiles and the corresponding bisectors resulting from the calculations of \ion{Fe}{i} 5576.09 \AA. The grey scale indicates the variation of the heliocentric angle from $\mu=1.0$ to $\mu=0.3$.}\label{velpan}
       
        \label{model}
\end{figure}

In \textbf{Paper I}, we extend the calibration method employed by \citet{2007langangen} for the \ion{C}{i} 5380 line. Snapshots from a 3D hydrodynamical simulation are used to compute synthetic profiles assuming Local Thermodynamic Equilibrium (LTE) (see Fig.~\ref{granmodel}). The convective shift of the spatially-averaged profile is measured from spectra computed with the numerical simulation. Our calculations are performed for eleven selected lines of interest for solar observers (listed in Table \ref{tablines}), over a range of heliocentric angles (distance from solar disc centre). The synthetic line profiles are provided in digital form to the community.
\begin{table}
\caption{Spectral lines used to create calibration data in Paper I.}             % title of Table
\label{tablines}      % is used to refer this table in the text
\centering                          % used for centering table
\begin{tabular}{l r l}        % centered columns (4 columns)
\hline\hline                 % inserts double horizontal lines
\ion{C}{i} & 5380.34 & forms in deep layers of the photosphere \\
\ion{Fe}{i} & 5250.21 & magnetometry \\
\ion{Fe}{i} & 5250.63 & magnetometry \\
\ion{Fe}{i} & 5576.09 & Doppler measurements\\
\ion{Fe}{i} & 6082.71 & abundance indicator \\
\ion{Fe}{i} & 6301.50 & magnetometry \\
\ion{Fe}{i} & 6302.49 & magnetometry \\
\ion{Fe}{i} & 7090.38 & Doppler measurements\\
\ion{Ca}{ii} & 8498.01 & chromospheric diagnostics\\
\ion{Ca}{ii} & 8542.05 & chromospheric diagnostics\\
\ion{Ca}{ii} & 8662.16 & chromospheric diagnostics \\
\hline                                   %inserts single line
\end{tabular}
\end{table}
This method assumes that 3D models can reproduce the correlation between brightness and Doppler shift of granulation in a statistical sense (see Fig. \ref{model}). 
The elemental abundances is used as a free parameters to achieve the best possible agreement between the synthetic profiles and the FTS atlas. The estimated parameters should not be regarded as abundances, as they also compensate for uncertainties in the atomic data, the LTE approximation used in the radiative transfer calculations, and other errors. The accuracy of our results is inferred from experiments carried out using the 3D models. From this and from observational tests, we estimate the results to have an accuracy of $~50~\mbox{m} \ \mbox{s}^{-1}$ at solar disk centre. 

Our results have been analyzed using bisectors. The bisector of a spectral line indicates the center of the profile as a function of intensity. Fig.~\ref{velpan} illustrate our results for the \ion{Fe}{i} 5576.09 \AA \ line. The line bisectors are shown along with the profiles at different heliocentric angles.  

In addition to providing calibration data, \textbf{Paper I} discusses the variations of the bisectors with disk position ($\mu$). This {\em limb effect} is found to mainly be caused by the 3D structure of the granulation, while the changing intensity-velocity correlation with height plays a minor role.

\chapter{Chromospheric diagnostics}\label{chrodiag}
\begin{figure}[]
      \centering
       \resizebox{\hsize}{!}{\includegraphics[trim=0 0 0 0.1cm]{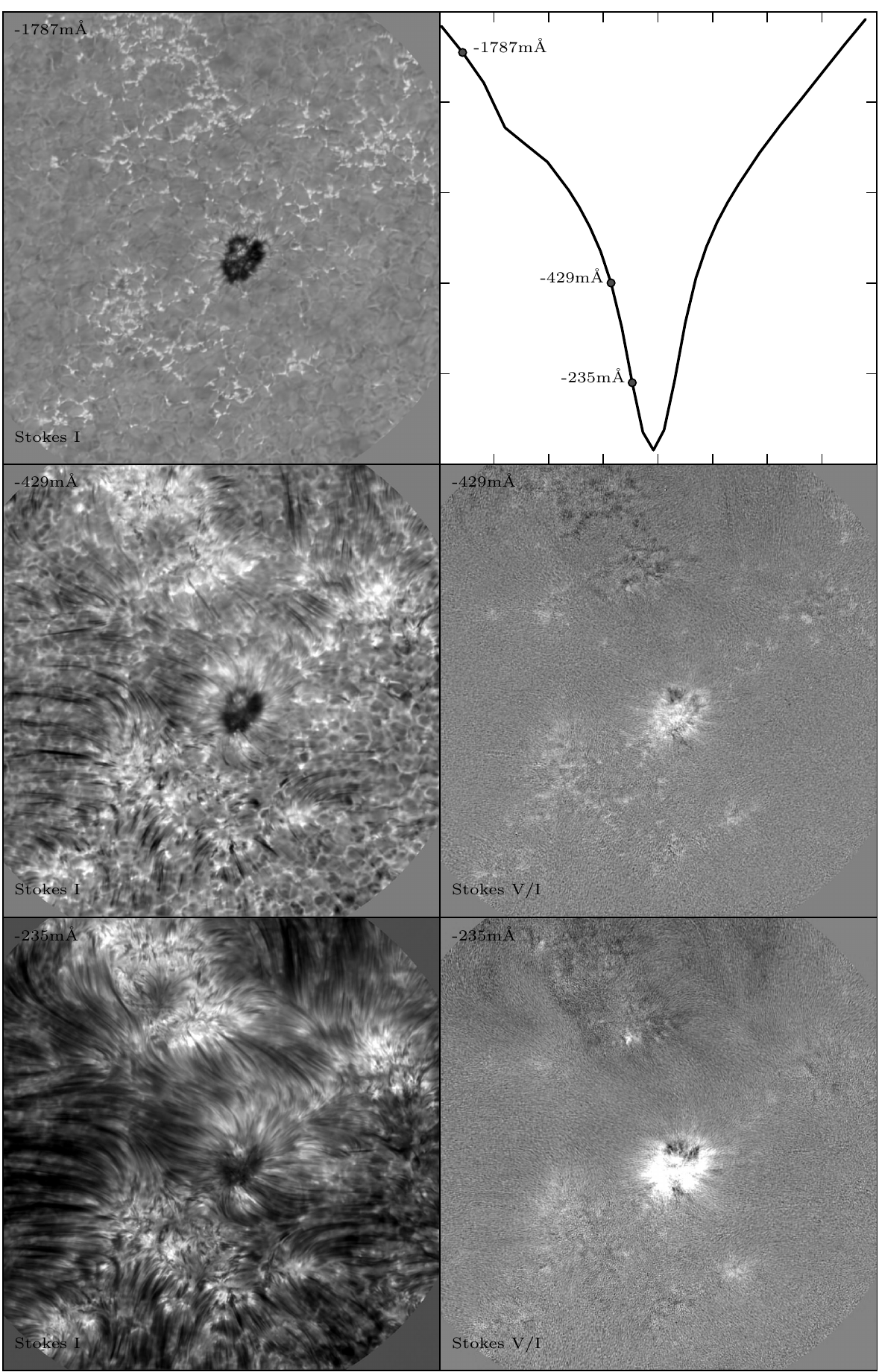}}
        \caption{
        Observations of \ion{Ca}{ii} 8542 \AA. The images show a transition from the photospheric wings of the line to the chromospheric core. The dataset was acquired by Luc Rouppe van der Voort (ITA-UiO) with SST/CRISP. 
       }
        \label{ca2sketch}
\end{figure}

It was mentioned in Chapter \ref{intro} that chromospheric observations are more difficult than those of the photosphere, especially when polarimetric measurements are involved. Spectral lines that are sensitive to the chromospheric range are usually very broad and only the core, where less photons are emitted, shows chromospheric features \citep{2008cauzzi}, as illustrated in Fig. \ref{ca2sketch} where the granulation present in the wings smoothly changes into a chromopheric landscape close to the core of the line. The lack of light, in combination with a broad profile and weaker magnetic fields than in the photosphere conspire to reduce the amplitude of Stokes $Q$, $U$ and $V$ profiles. 

The obvious solution to this problem would be to increase the exposure time of observations. However, the chromosphere is vigorously dynamic and long integration times usually translate into image smearing. The evolution time scale in the chromosphere can be estimated using an estimate of the Alfv\'en speed $v_A(B=100 \ \mbox{G}, z=1000 \ \mbox{km})\sim 10^5 \ \mbox{m} \ \mbox{s}^{-1} $ in the chromosphere \citep[see page 83,][]{1982priest}. In the case of the SST, the diffraction limit at 854.2 nm is $0."18$ which corresponds to 130 km on the surface of the Sun. Thus, the chromosphere cannot be assumed to be static for times longer than 1.3 seconds \citep[see][]{2006noort2}. 

Furthermore, this time scale also limits the spectral coverage of FPI observations, as only one wavelength can be observed at the time. It is normally assumed that the Sun does not change during a full scan of the line. If the the spectrum is sampled using a large number of frequency points, this assumption may not hold. Therefore, observing the chromosphere involves a trade-off between sensitivity, cadence and wavelength coverage. 

\section{Detectability of magnetic fields in the chromosphere}\label{detect}
During the past decade, the lines of the \ion{Ca}{ii} infrared triplet have been extensively used to diagnose the chromosphere \citep[see][and references therein]{2008langangen,2009leenaarts,2009cauzzi}, sometimes including polarization. Observational papers have usually studied cases with relatively strong magnetic field \citep{2000socas-navarro,2007pietarila,2010judge,2010delacruz}, whereas theoretical approaches have been restricted to 1D models \citep{2007pietarila2,2010manso}. 

\begin{figure}[]
      \centering
       \resizebox{0.9\hsize}{!}{\includegraphics[trim=0 0.7cm 0 0.1cm, clip]{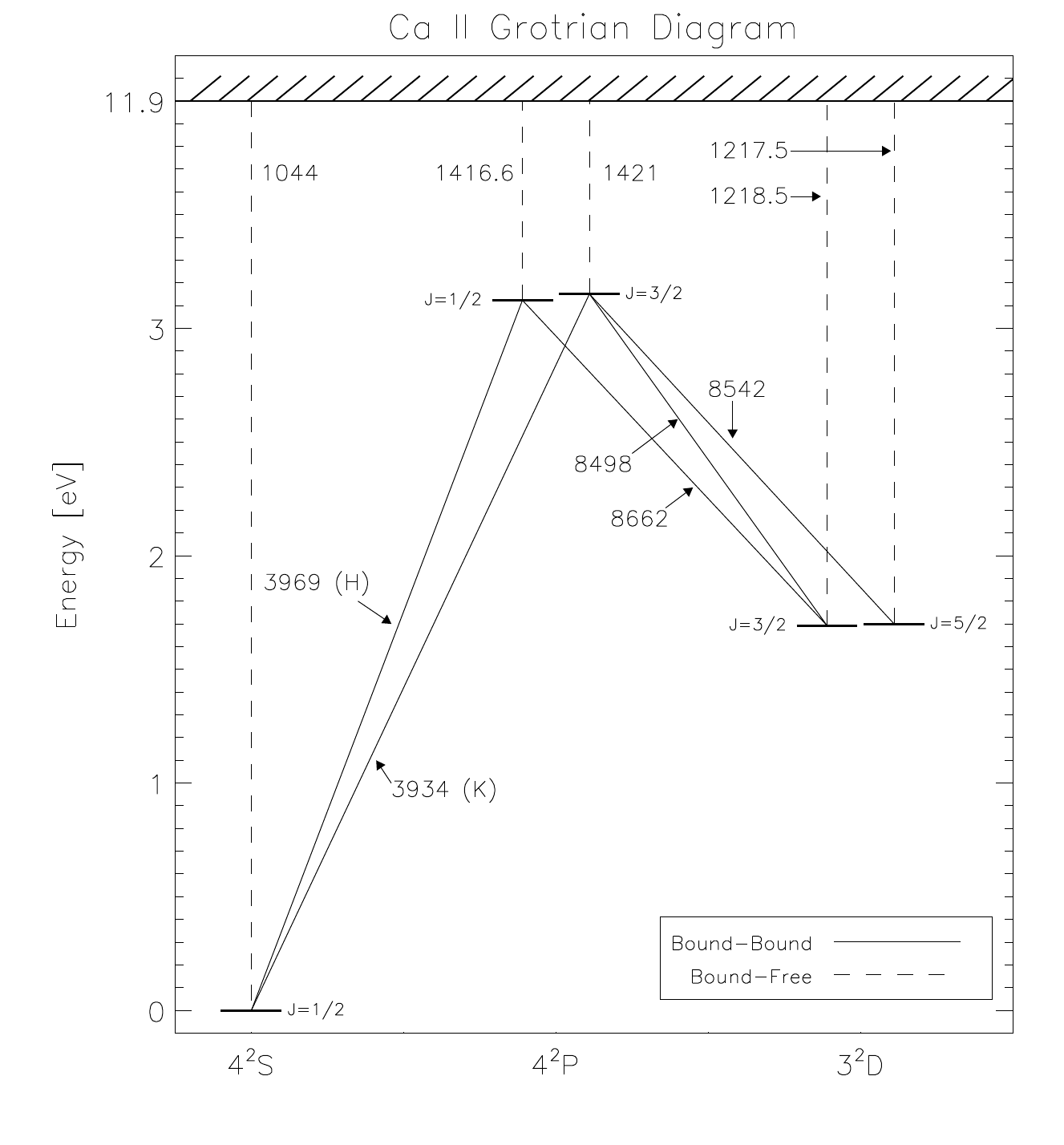}}
        \caption{
       \ion{Ca}{ii} atom model used to compute the calcium lines in \textbf{Paper I \& II}. Transitions between bound states are marked with solid lines, whereas transitions from a bound state to a free state are represented by dashed lines. The wavelengths of the transitions are given in \AA.
       }
        \label{ca2gro}
\end{figure}

In \textbf{Paper II} a snapshot from a realistic simulation of the chromosphere is used for the first time to compute synthetic full Stokes spectra. We use a simplified \ion{Ca}{ii} model atom that consists of 5 bound levels plus ionization continuum, which is illustrated in Fig. \ref{ca2gro}. The populations of the atom are computed in non-LTE evaluating the 3D radiation field as in \citet{2009leenaarts}. 

\begin{figure}[!h]
      \centering
       \resizebox{\hsize}{!}{\includegraphics[trim=0 0.cm 0 0.cm, clip]{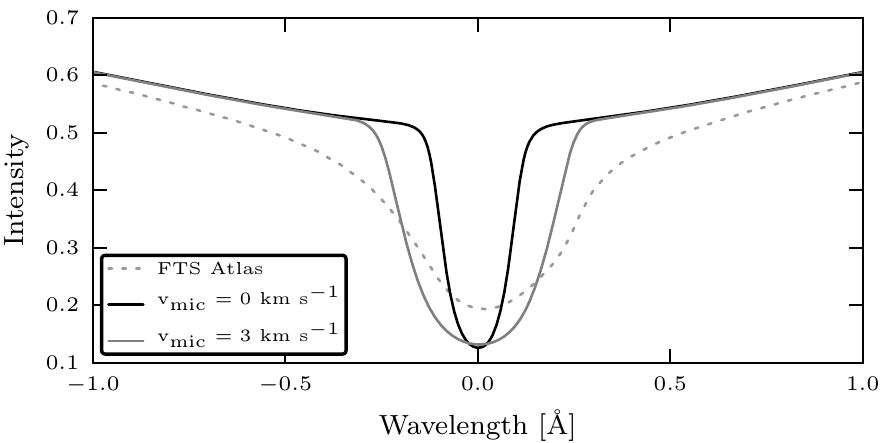}}

        \caption{
       Spatially-averaged spectrum from the 3D simulation computed without microturbulence (solid-black line) and with microturbulence (solid-grey line). The solar FTS atlas is plotted for comparison (dashed-line).
       }
        \label{mic}
\end{figure}
\begin{figure}[]
      \centering
       \resizebox{\hsize}{!}{\includegraphics[trim=0 0.6cm 0 0.cm, clip]{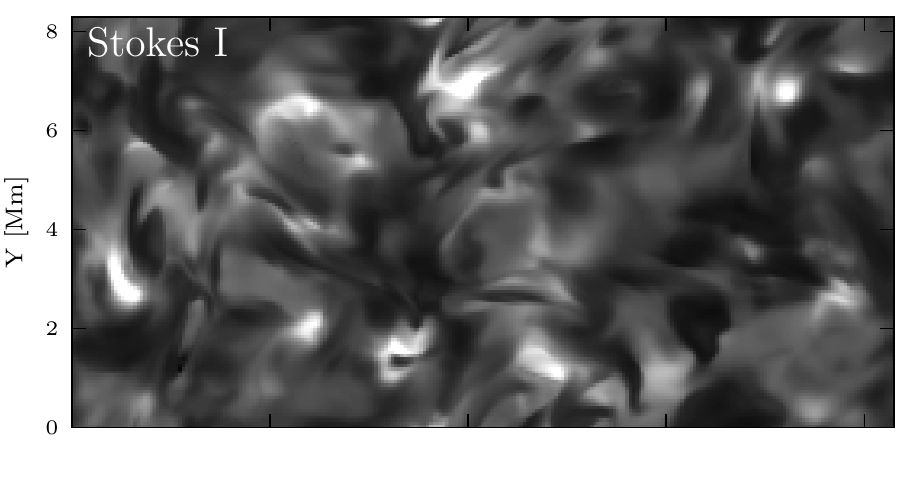}\includegraphics[trim=0.65cm 0.6cm 0 0.cm, clip]{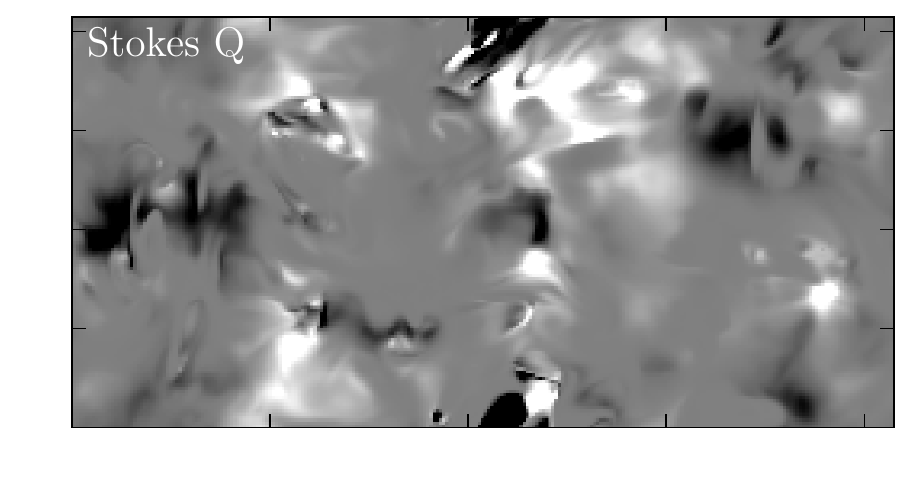}}
        \resizebox{\hsize}{!}{\includegraphics[trim=0 0.cm 0 0.cm, clip]{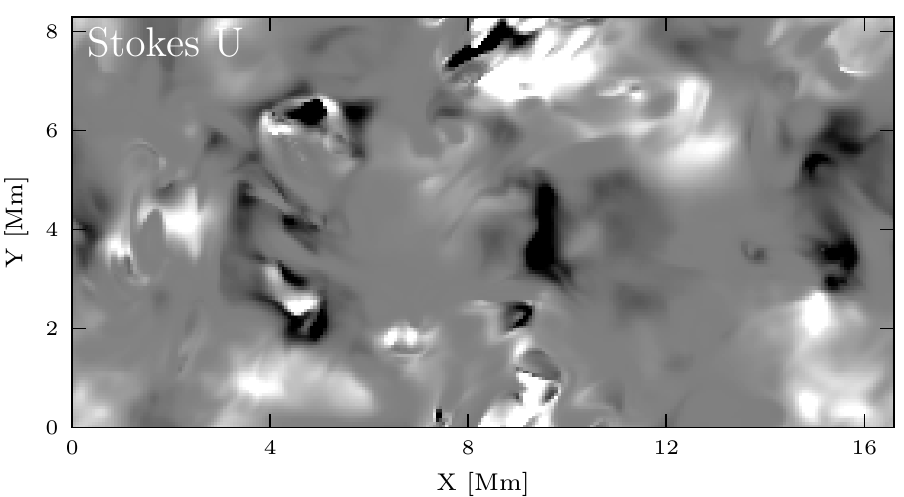}\includegraphics[trim=0.65cm 0.cm 0 0.cm, clip]{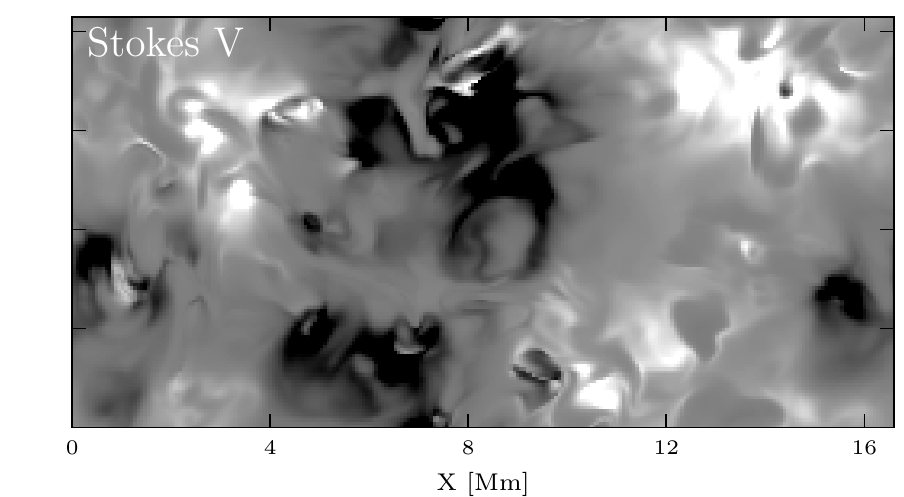}}
        %\resizebox{0.8\hsize}{!}{\includegraphics[trim=0 0.70cm 0 0.cm, clip]{figures/pan_stokesi.pdf}}
        %\resizebox{0.8\hsize}{!}{\includegraphics[trim=0 0.70cm 0 0.cm, clip]{figures/pan_stokesq.pdf}}
        %\resizebox{0.8\hsize}{!}{\includegraphics[trim=0 0.70cm 0 0.cm, clip]{figures/pan_stokesu.pdf}}
        %\resizebox{0.8\hsize}{!}{\includegraphics[trim=0 0.70cm 0 0.cm, clip]{figures/pan_stokesv.pdf}}
        \caption{
       Stokes images at $-75$ m\AA \  from the core of the line. The Stokes $Q$ and $U$ images are scaled between $\pm0.002$ and Stokes $V$ is scaled between $\pm 0.02$, relative to continuum intensity.
       }
        \label{ca2stokes}
\end{figure}
This study is partially motivated by the ongoing debate on requirements of the instrumentation needed to observe chromospheric polarization in the quiet Sun using the \ion{Ca}{ii} infrared triplet lines. We study the combined effect of spectral resolution and noise on our simulated observations of the chromosphere, considering the following:
\begin{itemize}
	\item All the polarization is due to the Zeeman effect. We neglect the Hanle effect, which depolarizes or polarizes the light depending on the scattering geometry and changes the ratio between $Q$ and $U$ \citep{2010manso}.
	\item The cores of our synthetic profiles are unrealistically narrow, probably because the model is missing small scale random motions. Conclusions based on these results would underestimate the effect of noise and overestimate the effects of instrumental degradation. Thus, we use microturbulence to broaden our profiles to the same width that is observed in spatially-resolved profiles. In Fig.~\ref{mic} the spatially-averaged spectrum from the 3D simulation with and without microturbulence are compared with the FTS atlas \citep[see][]{fts-atlas}.
	\item Instrumental degradation is described by a Gaussian point spread function that operates on the spectra. Additive random noise following a Gaussian distribution is introduced after the convolution with the instrumental profile.
\end{itemize}
Full Stokes monochromatic images computed  from the 3D simulations in the \ion{Ca}{ii}~8542~\AA \ line are shown in Fig.~\ref{ca2stokes}. The images have a lot of sharp features that are partially produced by Doppler shifts of the line. As the chromospheric core of the synthetic spectra is unrealistically narrow and strong, intensity variation from Doppler shifts are stronger than in reality.

Our results suggest that current FPI instruments are not sufficiently sensitive to detect circular polarization in the quiet Sun chromosphere using the \ion{Ca}{ii} 8542 \AA  \ line.

\section{Non-LTE inversions from a 3D MHD simulation}
Inversion codes have been extensively used to infer physical quantities from spectrometric and spectropolarimetric data. Inversions involve least-squares fits of the parameters of an atmospheric model, in order to reproduce observed profiles. In the photosphere,  LTE conditions or a Milne-Eddington atmosphere \citep{1992ruiz-cobo,2002bellot-rubio2} are often assumed to simplify and speed up the radiative transfer calculations. Interesting results have been achieved with Milne-Eddington inversions of the chromospheric \ion{He}{i} 10830 \AA \ lines \citep{2004lagg,2009asensio-ramos,2010kuckein}, which seem to form in the upper chromosphere \citep{2008centeno}. %The main limitation of these type of observations is that the limited spatial resolution of telescopes at $1~\mu\mbox{m}$ does not benefit from the impossibility of applying image reconstruction on slit-based data.

The work presented by \citet{2000socas-navarro0} demonstrated that non-LTE inversions of solar observations are possible. Along those lines, \citet{2007pietarila} carried out inversions of \ion{Ca}{ii} 8542 \AA \ data, to measure chromospheric quantities in quiet Sun. More recently,  \citet{2010delacruz} used the same scheme to carry out inversions on very high resolution observations of a sunspot showing umbral flashes.

\begin{figure}[]
      \centering
       \resizebox{\hsize}{!}{\includegraphics[]{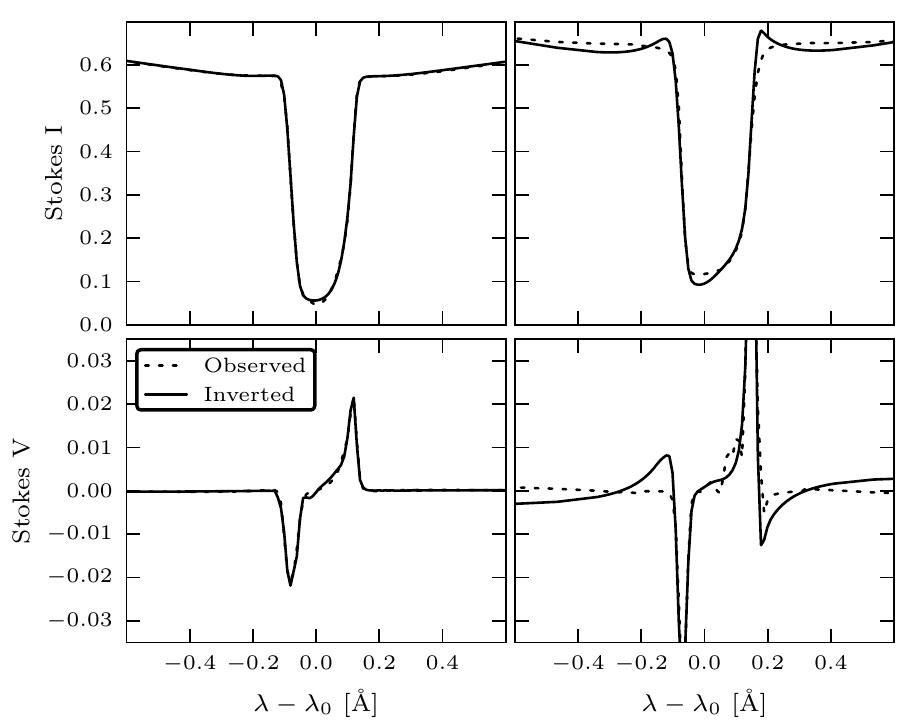}}
        \caption{
      Illustrative examples of the fits (solid-line) to the simulated observations. The left column is an example of a good fit, whereas the column on the right illustrates a poorer fit. The top row corresponds to Stokes $I$ and the bottom row to Stokes $V$.
      }
        \label{inv}
\end{figure}

However, it is hard to quantify how accurately inversions can provide chromospheric information, with the commonly used assumptions related to the radiative transfer calculations:

\begin{itemize}
	\item The populations of the levels of the atom are computed in non-LTE assuming plane-parallel geometry.
	\item Optionally, the computation of populations can be accelerated by neglecting the effects of the velocity field in the outcoming intensity. Thus, fewer azimuthal angles can be used to evaluate the radiation field. Under those conditions, the line profile becomes symmetric and only one half needs to be computed.
	\item The fitted model is assumed to be in hydrostatic equilibrium to impose consistency between temperature and density.
\end{itemize}
In \textbf{Paper II}, we use the synthetic observations described in Section \ref{detect}, without microturbulence, to test the Non-LTE Inversion Code based on the Lorien Engine (NICOLE) \citep{nicoleref}. The results of the inversion are compared with the quantities from the 3D simulation model. The inversions provide a good estimate of the chromospheric average line-of-sight velocity and magnetic field. 3D non-LTE effects could be affecting temperature, which presents less temperature contrast than the original model. Fig.~\ref{inv} shows two examples of fitted profiles from different pixels. The left panel corresponds to a good fit of the line, whereas the right panel shows a poor fit to the observed profiles. These failures originate the inversion noise that is mentioned in \textbf{Paper II}.
%\begin{figure}[]
%      \centering
%       \resizebox{0.9\hsize}{!}{\includegraphics[]{figures/inversion.pdf}}
%        \caption{
%       Flow of an \emph{Inversion}. A least-squares fitting scheme is used to infer the atmospheric parameters of a model. The algorithm minimizes the difference between observed and synthetic profiles that are computed using the atmospheric model.
%       }
%        \label{inv}
%\end{figure}
\section{Magnetic fields in chromospheric fibrils}
\begin{figure}[]
      \centering
       \resizebox{\hsize}{!}{\includegraphics[trim=0.1cm 0.3cm 0.45cm 0, clip]{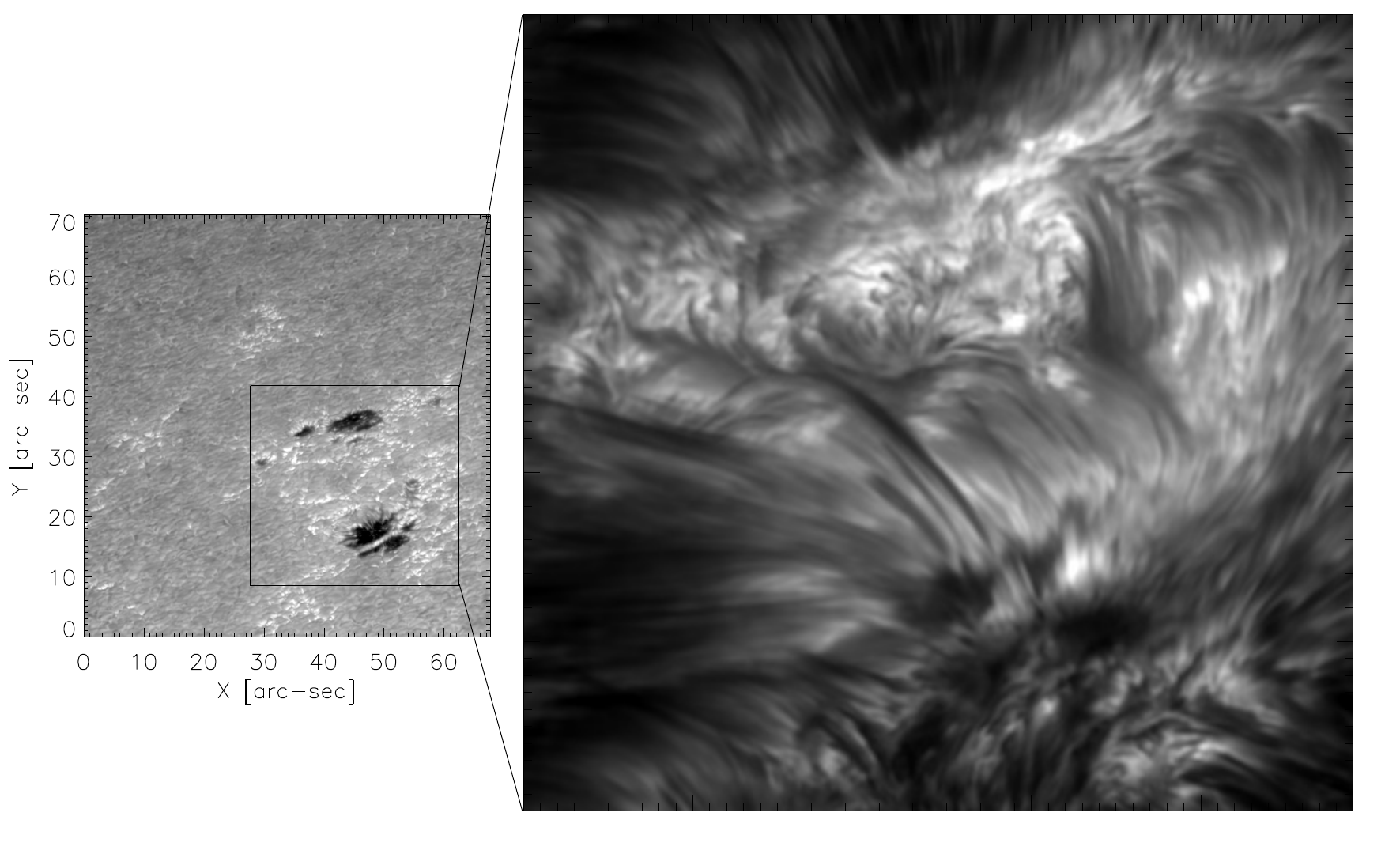}}
        \caption{
         Superpenumbral fibrils in the surrounding of two sunspots observed in \ion{Ca}{ii} 8542 \AA. The left panel shows a wideband image of the photosphere whereas the panel on the right corresponds to a narrowband image acquired at -161 m\AA  \ from the core of the line. Images acquired with the SST.
       }
        \label{fig:fibrils}
\end{figure}

It is widely assumed that fibrils outline chromospheric magnetic fields. Fibrils usually appear around facular regions in \ion{Ca}{ii} 8542 filtergrams \citep{2007rutten}, supporting the connection between fibrils and magnetic fields. The goal is to find  direct observational evidence of the alignment between magnetic fields and fibrils. We use two full Stokes datasets acquired in different telescopes with instruments of different type, to measure the orientation of magnetic fields in superpenumbral fibrils. The first dataset was acquired with SPINOR \citep{2006socas-navarro} at the Dunn Solar Telescope (DST), a slit based instrument that allows a large wavelength coverage at a spatial resolution of $0."6$. The second dataset is acquired with SST/CRISP at very high cadence achieving a spatial resolution of $0."18$ but with a limited spectral coverage (see Fig.~\ref{fig:fibrils}). In these datasets, the Stokes $Q$ and $U$ spectra are integrated along the length of the fibrils in order to improve the S/N ratio. The azimuthal direction of the magnetic field $(\chi)$ is calculated using the ratio between Stokes $Q$ and $U$.
\begin{equation}
%\tan (2\chi) = \frac{\int_0^\infty f(\lambda) U(\lambda)}{\int_0^\infty f(\lambda) Q(\lambda)},
\tan (2\chi) = \frac{ U}{Q} \label{azi}
\end{equation}
%Our measurements suggest that fibrils are mostly oriented along the magnetic field direction. However, interestingly, we find evidences of misalignment in some of our measurements. We cannot propose a scenario that explains this situation. \citet{2006judge} proposed that if $\beta<1$ in the chromosphere, then magnetic fields should be almost force-free showing smooth spatial variations. The fine structure seen in chromospheric observations should then primarily be produced by the thermodynamic properties of the gas. Our results could be compatible with this scenario. 
%However, we can assume a different situation where $\beta\sim~1$. Then, gas pressure  significantly contributes to the force balance and missalignments between fibrils and magnetic field could also happen.
Our measurements suggest that fibrils are mostly oriented along the magnetic field direction, however we find evidence of misalignment in some cases. This is both surprising, interesting, and hard to explain. \citet{2006judge} proposed that if $\beta < 1$ in the chromosphere, then magnetic fields should be almost force-free showing smooth spatial variations. The fine structure seen in chromospheric observations should then primarily be produced by the thermodynamic properties of the gas. Our results could be compatible with this scenario.

\chapter{Data collection and processing}\label{imred}
Some of the techniques that are described in this chapter are partially covered in \textbf{Paper IV}. However, the \emph{backscatter} problem (see \S \ref{backsc}) and the telescope polarization model (\S \ref{telmodel}) have not yet been described in separate publications, but are planned to  appear in a forthcoming paper \citep{2011delacruz}.

\section{The SST and CRISP}
\begin{figure}[]
      \centering
      \resizebox{0.8\hsize}{!}{\includegraphics[]{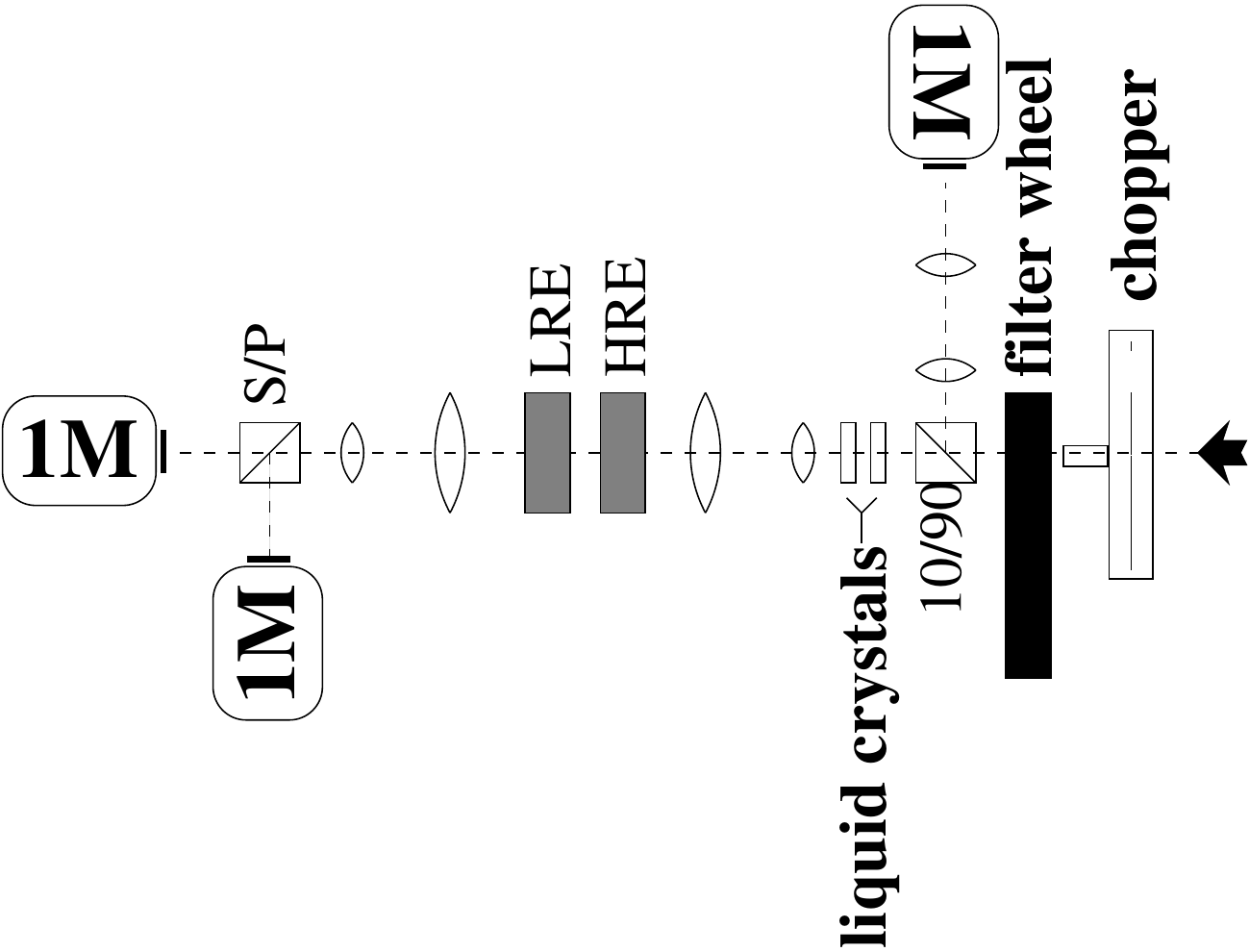}}
        \caption{
		Sketch of CRISP at the SST. The incoming light passes the chopper and the pre-filter that isolates the observed spectral line. A beam splitter separates 10 \% of the light to the WB camera. The rest of the light passes through the liquid crystals and the Fabry-P\'erot etalons (HRE and LRE). Finally a polarizing beam splitter separates the $p$ and $s$ components of the polarized light to different cameras (1M). 
       }
        \label{optical}
\end{figure}

The data presented in \textbf{Paper III} and \textbf{Paper IV} was acquired with the Swedish 1-m Solar Telescope (SST) \citep{2003scharmer}, located on the island of La Palma. Our narrow band data are acquired with the CRisp Imaging Spectropolarimeter \citep[CRISP,][]{2006scharmer} which is based on a Fabry-P\'erot interferometer that allow for narrow band observations at very high spatial resolution and cadence, providing spectral information at the same time. Atmospheric turbulence is compensated for with adaptive optics (AO), in order to improve image quality.
CRISP is mounted in the red beam of the SST  (see Fig.~\ref{optical}). The light that has been corrected by the AO system passes through the chopper and the pre-filter. Part of the light is reflected to the wideband camera. The other part is modulated with liquid crystals, producing linear combinations of the four stokes parameters. Afterwards, the light beam passes through the CRISP. The $p$ and $s$ polarizations are separated by a beam splitter into two beams that are recorded with separate cameras.
\section{Flat-fielding the data}\label{flats}
Science data taken with a CCD camera can be corrected for pixel-to-pixel inhomogeneities in the response of the camera. Normally, if the CCD is illuminated with a flat and homogeneous light source, intensity variations are mostly produced by pixel-to-pixel sensitivity variations, dirt and fringes. Thus, flat-field calibration images can be acquired to correct for these intensity variations. 

A particular problem arises by the presence of time-dependent telescope polarization in the data. Normally, the flat-field images are taken at a different time than the science data. Thus, the amount of polarization introduced by the telescope can differ significantly between science and flat field data. In principle this should not be a problem if our cameras could detect Stokes parameters directly. Instead, four linear combinations of Stokes $I$, $Q$, $U$ and $V$ are acquired. If seeing were not present, the demodulation of the data could be carried out directly. However, in our case image restoration needs to be done in order to remove residual effects of seeing, not fully compensated for by the AO. As the image reconstruction is done with modulated data, artifacts appear if the flats and science data are not taken at similar times. 

\begin{figure}[]
      \centering
      \resizebox{1.0\hsize}{!}{\includegraphics[]{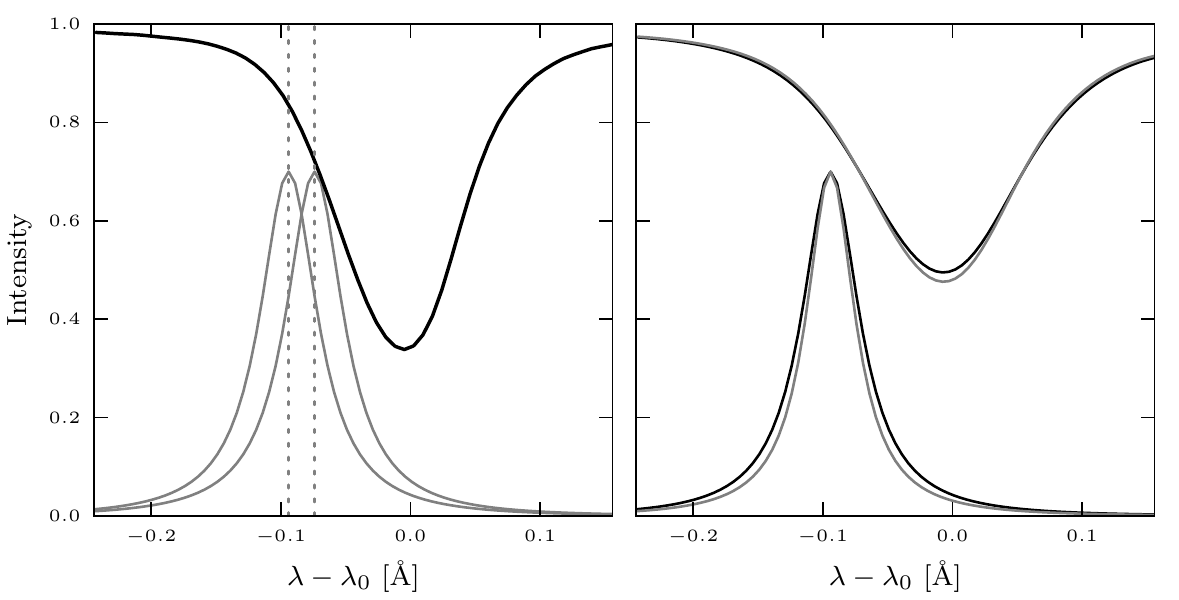}}
        \caption{
		\emph{Left:} A theoretical profile of CRISP is plotted with a grey line at two spectral locations. This example illustrates how a shift of the instrumental profile leads to a variation of the intensity on the flanks of a spectral line. The profile has been scaled to the intensity of the line at the location where it is centered. \emph{Right:} Intensity variations produced by reflectivity errors. The line has been convolved with the instrumental profile of the same color.
       }
        \label{cav_err}
\end{figure}

Additionally, flat-fielding narrow-band images is more complicated than flat-fielding wideband images. In the case of CRISP,  inhomogeneities on the surface of the FPI etalons produce field-dependent wavelength shifts of the transmission profile of the instrument. These are called cavity errors because they introduce variations in the FPI cavities that define the wavelength. The combined effect of cavity errors and the presence of a spectral line, produce field-dependent intensity variations purely introduced by the slope of the spectral line. At the same time, variations in the reflectivity of the etalons across the field-of-view translate into minor variations of the width of the instrumental profile, and therefore also to overall transmission variations. This effect is much smaller than intensity fluctuations produced by cavity errors. Fig. \ref{cav_err} illustrates these two effects on the \ion{Fe}{i} 6302 \AA \ line.

In \textbf{Paper IV} we address the flat-fielding problems produced by time varying telescope polarization and by cavity/reflectivity errors. We propose a method to flat-field polarimetric data affected by telescope polarization. A numerical scheme is used to model and remove the fingerprints of the spectral line from our flat-field data. 

%\begin{enumerate}
%	\item \label{li-re} A flat-field correction is applied to science images using flats that contain cavity errors. This partially wrong flat-field correction produced a smooth image that is suitable for the image reconstruction step that is carried out with MOMFBD \citep{2005noort}.
%	\item The flat-field data are demodulated and compensated for telescope polarization, producing a Stokes I flat at each observed wavelength. The Stokes I flats contains intensity fluctuations across the field of view that are produced by the combined presence of cavity errors and a spectral line.
%	\item  For each pixel, the parameters of a model are fitted using a Levenberg-Marquardt algorithm in order to reproduce the observed profile. The model accounts for global gain factors, cavity errors, reflectivity errors and pre-filter variations across the field of view. Remaining differences between the fitted spectrum and the observed one are considered genuine wavelength dependent gain factors.
%	\item New Stokes I flats are computed using the parameters of the model. The signature of the spectral line is not included in these new flats.
%	\item The reconstructed data from step \ref{li-re} are demodulated using a matrix that contains the the ratio between the flat used in the reconstruction and the new correct flat.
%\end{enumerate} 
%The demodulation matrix  is calibrated on each pixel using calibration optics. However, the coupling between camera pixels and image pixels is lost in the reconstruction step. 

\section{The backscatter problem}\label{backsc}
The content of this section only applies to observations carried out at infrared wavelengths, and it was motivated by our first observations in \ion{Ca}{ii} 8542 \AA. The methods described here were developed together with Michiel van Noort and it is the main result of my first campaign with CRISP in 2008.
\subsection*{The problem}\label{bprob}
\begin{figure}[]
      \centering
      \resizebox{1.0\hsize}{!}{\includegraphics[trim=0.2cm 0 0.1cm 0, clip]{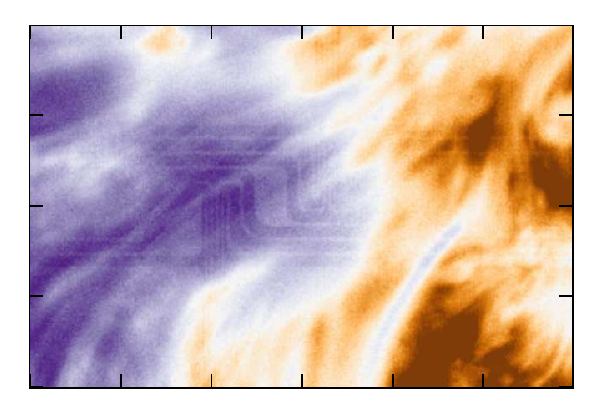}\includegraphics[trim=0.2cm 0 0.1cm 0.cm, clip]{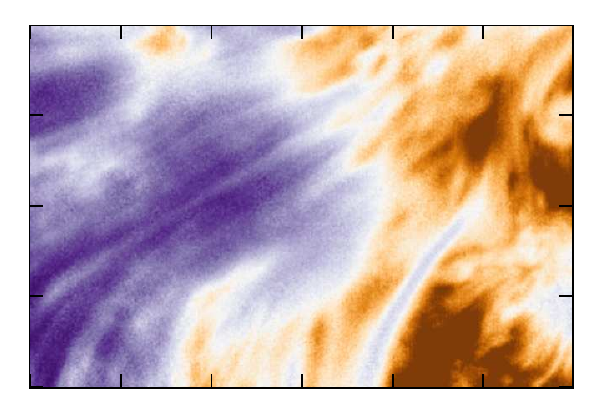}}
        \caption{
		\emph{Left:} Subfield of an image that has been traditionally flat-fielded, showing the circuit-like pattern. \emph{Right:} The same subfield corrected using the numerical scheme that is proposed in this section. This colormap is chosen in order to enhance the contrast of fibrils and the electronic circuit.  
       }
        \label{bsc_comp}
\end{figure}

The CRISP acquisition system includes three back-illuminated CCD cameras (Sarnoff) that can record 35 frames per second. The quantum efficiency of these cameras decreases towards long wavelengths. Above 700 nm the CCD becomes semi-transparent, letting part of the light to pass through. Furthermore, the images show a circuit-like pattern that cannot be removed by traditional dark-field and flat-field corrections, as illustrated in Fig. \ref{bsc_comp}.\\
Examination of pinhole data and flat-fielded data shows that:
\begin{enumerate}
	\item There is a diffuse additive stray-light contribution in the whole image.
	\item The stray-light contribution is much smaller in the circuit pattern and the gain appears to be enhanced.
\end{enumerate}
\subsection*{The model}\label{bmod}

\begin{figure}[]
      \centering
      \resizebox{0.8\hsize}{!}{\includegraphics[]{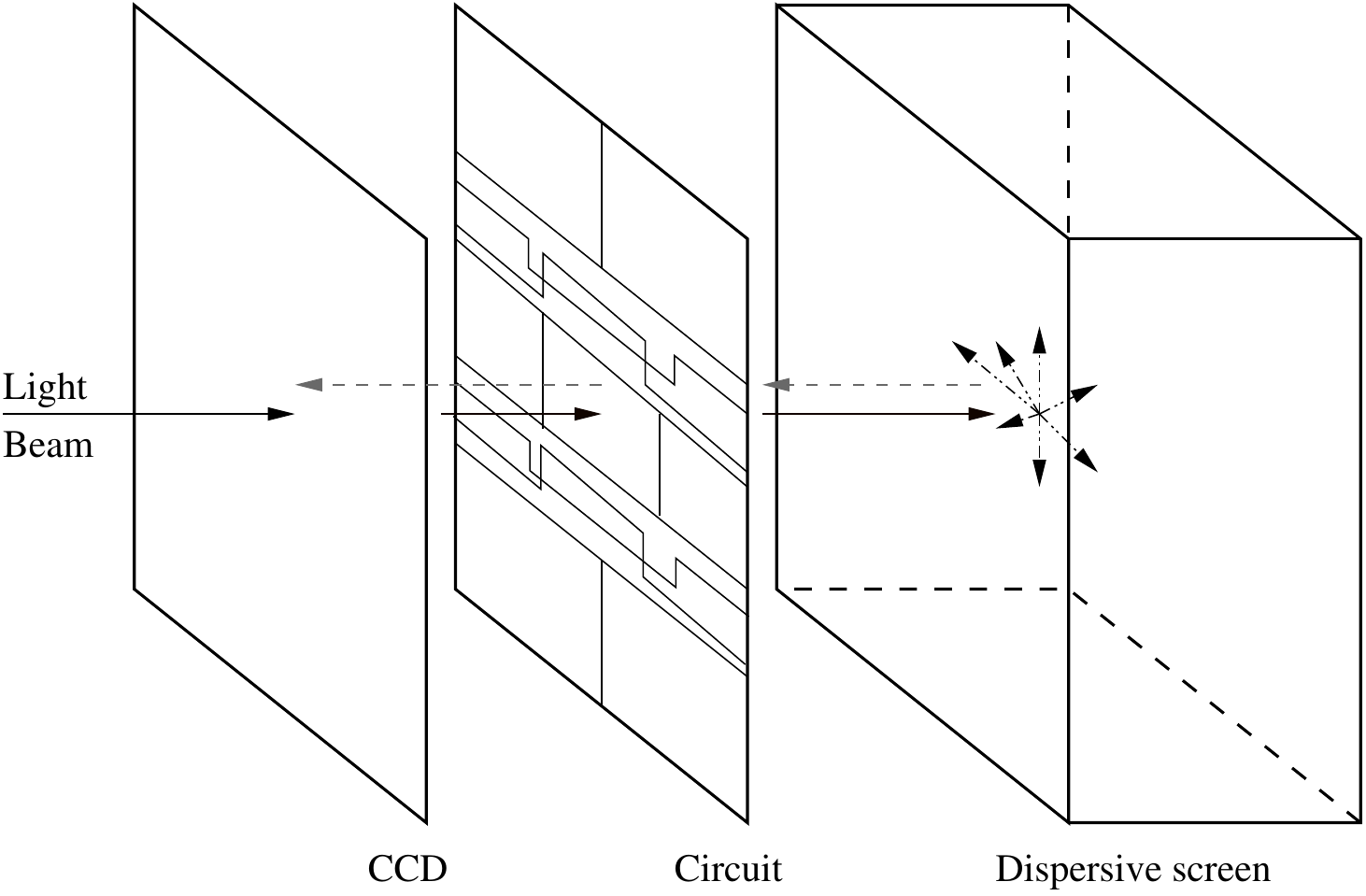}}
        \caption{
		Cartoon showing the conceptual model of the camera. The dark arrows indicate the incoming light beam from the telescope and the gray arrows represent the back-scattered light that returns to the CCD. 
       }
        \label{bscatter}
\end{figure}

These problems can be explained by a semi-transparent CCD with a diffusive medium behind it, combined with an electronic circuit located right behind the CCD that is partially reflecting and therefore also less transparent to the scattered light. Fig.~\ref{bscatter} represents the simplified structure of the camera. Under normal conditions an image recorded with a CCD, $I_o$, can be described in terms of the dark-field $D$, the gain factor $G_f$ and the real image $I_r$:
\begin{equation}
	I_o=D+(G_f \cdot I_r)\label{nimg}
\end{equation}
However, in the infrared we need a more complicated model:
\begin{equation}
	I_o = D +f(1 - f )[(G_b G_f I_r ) * P] G_b + f G_f I_r \label{sfull}
	%I_o=D+G_b[(G_b G_f I_r)\otimes P] (1-f)+ G_f I_r\label{sfull}
\end{equation}
where $f$ represents the overall fraction of light absorbed by the CCD, $G_b$ is the gain for light illuminating the CCD from the back which should account for the electronic circuit pattern. In the following, we refer to $G_b$ as \emph{backgain}. $P$ is a Point Spread Function (PSF) that describes the scattering properties of the dispersive screen. In the backscatter term we assume that $(1-f)G_fI_rG_b$ is transmitted to the diffusive medium where it is scattered. A fraction of the scattered light returns to the CCD, passing again through the circuit. The cartoon in  Fig.~\ref{bscatter} shows a schematic representation of the structure of the camera and the path followed by the light beam.

\subsection*{The numerical approach}\label{bnum}

\begin{figure}[]
      \centering
      \resizebox{0.9\hsize}{!}{\includegraphics[]{figures/bars_00}\includegraphics[]{figures/bars_01}}
     \resizebox{0.9\hsize}{!}{\includegraphics[]{figures/bars_02}\includegraphics[]{figures/bars_03}}
    % 
      %\resizebox{1.0\hsize}{!}{\includegraphics[]{figures/bars_00}\includegraphics[]{figures/bars_01}\includegraphics[]{figures/bars_02}\includegraphics[]{figures/bars_03}}
        \caption{
		Calibration images used to infer the scattering PSF and the backgain of the camera. The bars are displaced image-to-image so every pixel is covered at least once by the black bars. The images are shown with logarithmic intensity scale.
       }
        \label{bbars}
              \resizebox{1.0\hsize}{!}{\includegraphics[]{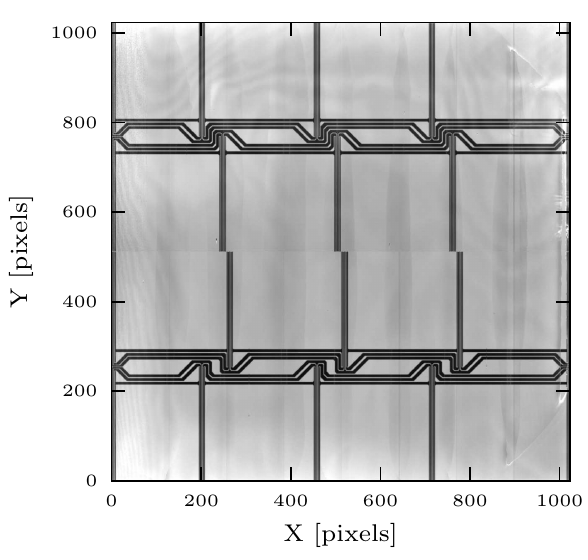}\includegraphics[trim=0.2cm 0 0.8cm 0,clip]{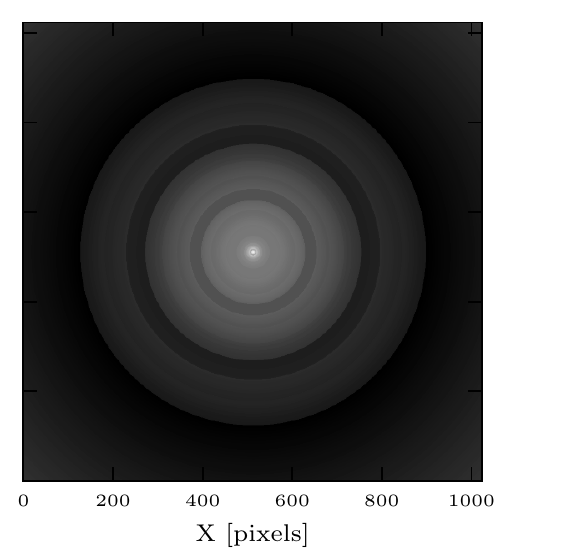}}
     % \resizebox{1.0\hsize}{!}{\includegraphics[]{figures/back_line.pdf}\includegraphics[]{figures/bpsf.pdf}}
        \caption{
		\emph{Left:} Fitted back-gain for one of the cameras. \emph{Right:} Circular-symmetric PSF obtained from our fitting routine. The PSF is displayed using a logarithmic scaling. 
       }
        \label{back_fit}
\end{figure}

In order to obtain the real intensity $I_r$, the PSF $P$, the back-gain $G_b$ and the front-gain $G_f$ must be known. The transparency factor is assumed to be smooth because the properties of the dispersive screen seem to be homogeneous across the field-of-view. This allows us to include $f$ in the front and the back gain factors
$$G_{b^\prime}=\sqrt{1-f} \ G_b$$
$$G_{f^\prime}=f G_f,$$
so Eq. \ref{sfull} becomes:
\begin{equation}
	I_o=D+G_{b^\prime}[(G_{b^\prime} G_{f^\prime} I_r)* P] + G_{f^\prime} I_r\label{sfullsim}
\end{equation}
This problem is linear and invertible, but a direct inversion would be expensive, given the dimensions of the problem. A numerical approach can be used to iteratively solve the problem. We define
\begin{equation}
\hat{J} = (G_{b^\prime} G_{f^\prime} I_r)* P
\end{equation}
In the first iteration, we initialize $\hat{J}$ assuming that the smearing caused by $P$ is so large that $\hat{I}$ can be approximated by the \textit{average} value of the observed image multiplied with the back gain. The back gain $G_{b^\prime}$ is assumed to be 1 for every pixel in the first iteration. Furthermore, we assume that the product $G_{f^\prime}I_r$ can be estimated from Eq. \ref{nimg} ignoring back scattering, i.e., $G_{f^\prime}I_r \approx I_o -D$. These values are of the same order of magnitude as the final solution, and therefore correspond to  a reasonable choice of initialization.
\begin{equation}
\hat{J}_0 \approx \langle G_{b^\prime} G_{f^\prime} I_r \rangle \approx \langle G_{b^\prime} (I_o - D) \rangle,
\end{equation}
where the initial guess of the PSF $P$ represents an angular average obtained from a pinhole image. The small diameter of the pinhole only allows to estimate accurately the central part of the PSF. The wings of our initial guess are extrapolated using a power-law. 
The estimate of $G_{f^\prime} I_r$ is then
\begin{equation}
G_{f^\prime} I_r = I_o -D -G_{b^\prime} \hat{J}
\end{equation}
which can be used to compute a new estimate $\hat{J}$. This procedure is iterated, until $G_{f^\prime} I_r$ and $I_o$ are consistent.
This new value of $\hat{J}$ is used to improve our estimate of $G_{b^\prime}$, by applying the same procedure to images that contain parts  being physically masked ($I_r = 0$), as in Fig. \ref{bbars}. 
In the masked parts where $I_r\equiv 0$ we have,
\begin{equation}
I_o - D - G_{b^\prime} \hat{J} = 0,
\end{equation}
so the back-gain can be computed directly:
\begin{equation}
G_{b^\prime}  = \frac{I_o-D}{\hat{J}}.\label{backg}
\end{equation}
Thus, every pixel must have been covered by the mask at least once in a calibration image in order to allow the calculation of the back-gain. With the new $G_{b^\prime}$ we can recompute a new estimate of $G_{f^\prime} I_r$.

We now need to specify a measure that describes how well the data are fitted by the estimate of $P$ and $G_{b^\prime}(P)$. Since both $G_{f^\prime} I_r$ and $G_{b^\prime}$  are computed based on self-consistency, we iteratively need to fit only the parameters of the PSF. We assume that the PSF is circular-symmetric and apply corrections to the PSF at \emph{node} points placed along the radius.

We use Brent's Method described by \citet{2002nr} to minimize our fitness function. This algorithm does not require the computation of derivatives with respect to the free parameters of the problem. When the opaque bars block a region of the CCD, an estimate of the back gain can be calculated for a given PSF according to Eq.~\ref{backg}. In our calibration data, the four bars of width $L$ are displaced $0.5 \ L$ from one image to the next (see Fig.~\ref{bbars}). This overlapping provides two different measurements of the back gain on each region of the CCD. However, as the location of the bars changes on each image, the scattered light contribution is different for each of these measurements of the back gain. Our fitness function minimizes the difference between these two measurements of the back gain. 

Having thus obtained the PSF $P$ and the back-gain $G_{b^\prime}$, we obtain $G_{f^\prime}$ by recording conventional flats and assuming $I_r$ is a constant in order to obtain $G_{f^\prime}$ from Eq.~\ref{sfullsim}.

\subsection*{Results}\label{backres}
The  numerical scheme  described in \S \ref{bnum} produces the backgain ($G_{b^\prime}$) and an approximate PSF ($P$) that describes the scattering problem. The results are illustrated in Fig. \ref{back_fit}. The  back gain image shows a background with vertical dark areas in the extended gaps where the circuit pattern is not present. Those hollows probably indicate that the PSF is not perfect, an expected result given the assumptions imposed on the shape of the PSF: it is constant over the entire field of view, and only radial variations are allowed. The PSF has extended wings that complicate the convergence of the solution, given the nature of our calibration data: the spacing between \emph{bars} in the horizontal direction is a limiting factor to constrain such extended wings.

The flat-fielding method described in the previous sections has been implemented in the image reconstruction code MOMFBD \citep{2005noort} and is only used when a back-gain and a PSF are provided. Eq. \ref{sfullsim} is reordered to correct the science data.
\begin{equation}
G_f I_r\label{sfullsim} = I_o - G_{b^\prime}[(G_{b^\prime} G_{f^\prime} I_r)* P]  - D \label{clean}
\end{equation}
Since $G_{f^\prime}I_r$ appears in both sides of the equation, some iterations are needed to estimate the term $G_{b^\prime}[(G_{b^\prime} G_{f^\prime} I_r)* P]$, which is a computationally expensive process if thousands of images are flat-fielded. 

\begin{figure}[]
      \centering
      \resizebox{1.0\hsize}{!}{\includegraphics[]{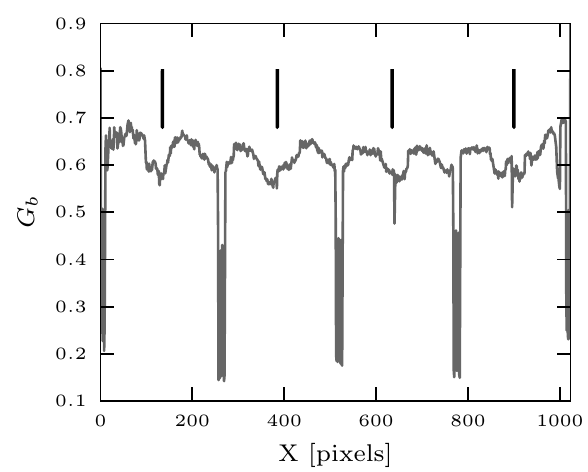}\includegraphics[]{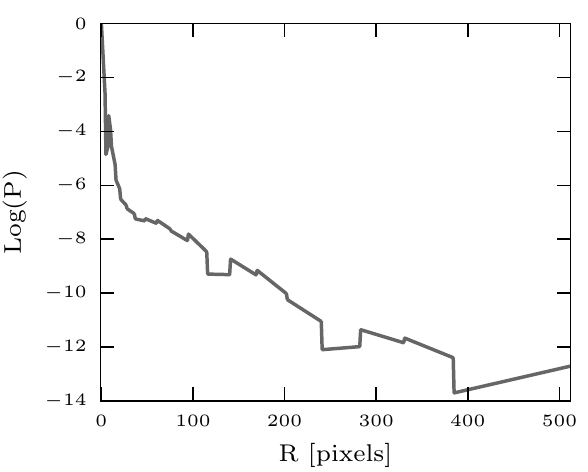}}
        \caption{
		\emph{Left:} Values of the back gain along the $x-$axis at $Y=450$. The black markers indicate the location of the hollows resulting from the limited accuracy of the PSF. \emph{Right:} The radial variation of $P$, represented in logarithmic scale.
       }
        \label{back_lin}
\end{figure}

In Fig. \ref{bsc_comp} we show a frame that has been traditionally flat-fielded (left panel) and the same image flat-fielded according to Eq.~\ref{clean}. The method described in the present work removes the electronic circuit pattern from the images. The assumption imposed on the PSF allows us to estimate the back gain with limited accuracy, as is obvious from the Fig.~\ref{back_lin}. The uncertainties present in the PSF and the back gain are likely to affect the contrast of the corrected images.

\newpage
\section{Telescope polarization model at 854.2 nm}\label{telmodel}
\begin{figure}[]
      \centering
      \resizebox{1.0\hsize}{!}{\includegraphics[]{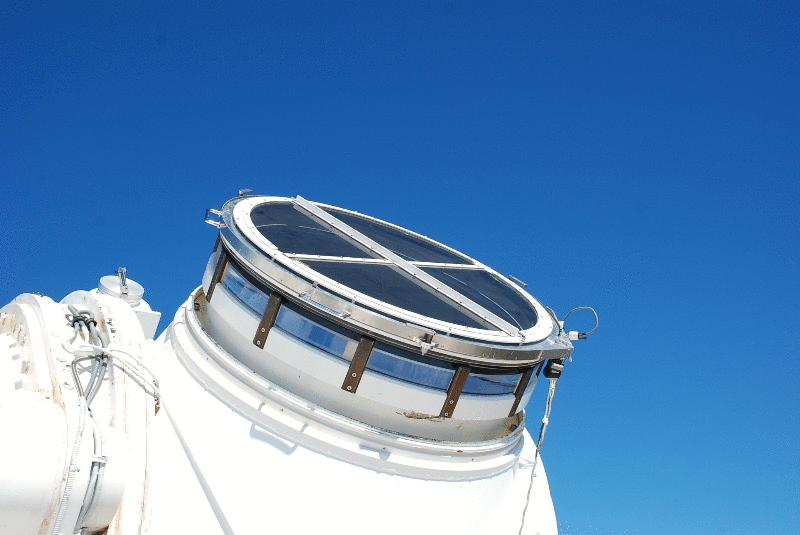}}
        \caption{
		The 1-m polarizer mounted on the entrance lens of the SST. Calibration datasets were acquired during entire days to characterize the polarimetric properties of the telescope.
       }
        \label{1mp}
\end{figure}

The turret of the SST contains optical elements that polarize the incoming light. \citet{2005selbing} studied the polarizing properties of the telescope at 630.2 nm and proposed a theoretical model to characterize its temporal variation. Calibration images have been taken using a 1-m polarizer mounted at the entrance lens of the SST (see Fig. \ref{1mp}). The polarizer rotates $360^{\circ}$ in steps of $5^\circ$ and several frames are acquired on each polarizer angle. Data were adquired during the whole day. These data were used to determine the parameters of the model proposed by \citet{2005selbing} at 854.2 nm. 
\subsection*{Telescope model}
Each polarizing optical element of the telescope is represented with a Mueller matrix. Mueller matrices of mirrors are noted with $M$ and have two free parameters. Assuming a wave that propagates along the $z$ axis and oscillates in the $x-y$ plane, the parameters are the de-attenuation term $(R)$ between the $xy$ components of the electromagnetic wave and the phase retardance ($\delta$) produced by the mirror \citep[see][]{2005selbing}. This form of M assumes that Q is perpendicular to the plane of incidence. In the more general case, $M$ can be written as,
$$
	M(R,\delta)  =  {\footnotesize\left( \begin{array}{c c c c}
		\frac{1}{2}+\frac{1}{2}R	&		\frac{1}{2}-\frac{1}{2}R	&	0		&	0 \\
		\frac{1}{2}-\frac{1}{2}R	&		\frac{1}{2}+\frac{1}{2}R	&	0		&	0 \\
		0		& 0	&	-\sqrt{R}\cos (\frac{\pi}{180}\delta)	&	-\sqrt{R}\sin(\frac{\pi}{180}\delta)	 \\
		0		& 0	&	\sqrt{R}\sin (\frac{\pi}{180}\delta)	&	-\sqrt{R}\cos(\frac{\pi}{180}\delta)	 \\
	\end{array}\right)}
$$
$R(\alpha)$ corresponds to a rotation to a new coordinate  frame,  rotated an angle $\alpha$. 
$$
	R(\alpha) = {\footnotesize \left( \begin{array}{c c c c}
		1		&	0		&	0		&	0 \\
		0		&	\cos(\frac{\pi}{90}\alpha) &	\sin (\frac{\pi}{90}\alpha)	&	0 \\
		0		&	-\sin(\frac{\pi}{90}\alpha)	&	\cos (\frac{\pi}{90}\alpha)	&	0 \\
		0		& 0	&	0	&	1
	\end{array} \right)}
$$
 $L$ represents the entrance Lens. The form of the lens matrix represents a composite of random retarders, so it modulates the light without (de)polarizing it. The reference for Q is aligned with the 1m linear polarizer axis and the values of the matrix are measured in that frame.

$$
	L =  {\footnotesize\left( \begin{array}{c c c c}
		1		&	0		&	0		&	0 \\
		0		&	A		&	B		&-C \\
		0		&	B		&	D		&E\\
		0		& 	C		& -E	&	(A+D-1)
	\end{array} \right)}
$$
%The lens matrix does has an arbitrary reference for Q, but it is not changed by the lens. When the lens weakly depolarizes the incident light, $A,D\approx1$ and $B,C,E\approx0$, very close to the identity matrix.
The model is built using the Mueller matrix of each polarizing element in the telescope (Eq. \ref{telmod}), as a function of the azimuth $(\varphi)$ and elevation $(\theta)$ angles of the Sun at the time of the observation. We have included the conversion factor from degrees to radians in the matrices, thus all the angles are given in degrees.
\begin{equation}
	M_{tel}\left(\theta,\varphi\right) =  R_{f+} \cdot M_{s}\cdot M_{f} \cdot R_{f-}\cdot R_{az}\cdot M_{az} \cdot R_{el}\cdot M_{el} \cdot L \label{telmod}
\end{equation}
$$
	L = {\footnotesize \left( \begin{array}{c c c c}
		1		&	0		&	0		&	0 \\
		0		&	c_0	&	c_1	&	-c_2 \\
		0		&	c_1	&	c_3	&	c_4 \\
		0		& c_2	&	-c_4	&	c_0+c_3-1
	\end{array} \right)}
$$
$$
	R_{el} = {\footnotesize\left( \begin{array}{c c c c}
		1		&	0		&	0		&	0 \\
		0		&	\cos (\pi(1+\frac{\theta}{90})) &	\sin (\pi(1+\frac{\theta}{90}))	&	0 \\
		0		&	-\sin (\pi(1+\frac{\theta}{90}))	&	\cos (\pi(1+\frac{\theta}{90}))	&	0 \\
		0		& 0	&	0	&	1
	\end{array} \right)}
$$
$$
	M_{az}  = {\footnotesize\left( \begin{array}{c c c c}
		\frac{1}{2}+\frac{1}{2}c_5	&		\frac{1}{2}-\frac{1}{2}c_5	&	0		&	0 \\
		\frac{1}{2}-\frac{1}{2}c_5	&		\frac{1}{2}+\frac{1}{2}c_5	&	0		&	0 \\
		0		& 0	&	-\sqrt{c_5}\cos (\frac{\pi}{180}c_6)	&	-\sqrt{c_5}\sin(\frac{\pi}{180}c_6)	 \\
		0		& 0	&	\sqrt{c_5}\sin (\frac{\pi}{180}c_6)	&	-\sqrt{c_5}\cos(\frac{\pi}{180}c_6)	 \\
	\end{array} \right) = M_{el}}
$$
$$
	R_{az} = {\footnotesize\left( \begin{array}{c c c c}
		1		&	0		&	0		&	0 \\
		0		&	\cos(\frac{\pi}{90}\varphi) &	\sin (\frac{\pi}{90}\varphi)	&	0 \\
		0		&	-\sin(\frac{\pi}{90}\varphi)	&	\cos (\frac{\pi}{90}\varphi)	&	0 \\
		0		& 0	&	0	&	1
	\end{array} \right)}
$$
$$
	R_{f+} = {\footnotesize\left( \begin{array}{c c c c}
		1		&	0		&	0		&	0 \\
		0		&	\cos (\frac{\pi}{90}c_{11}) &	\sin(\frac{\pi}{90}c_{11})	&	0 \\
		0		&	-\sin (\frac{\pi}{90}c_{11})	&	\cos(\frac{\pi}{90}c_{11})	&	0 \\
		0		& 0	&	0	&	1
	\end{array} \right)}
$$
$$
	R_{f-} ={\footnotesize \left( \begin{array}{c c c c}
		1		&	0		&	0		&	0 \\
		0		&	\cos (-\frac{\pi}{90}c_{11}) &	-\sin (-\frac{\pi}{90}c_{11})	&	0 \\
		0		&	\sin (-\frac{\pi}{90}c_{11})	&	\cos (-\frac{\pi}{90}c_{11})	&	0 \\
		0		& 0	&	0	&	1
	\end{array} \right)}
$$
$$
	M_{f} = {\footnotesize\left( \begin{array}{c c c c}
		\frac{1}{2}+\frac{1}{2}c_7	&		\frac{1}{2}-\frac{1}{2}c_7	&	0		&	0 \\
		\frac{1}{2}-\frac{1}{2}c_7	&		\frac{1}{2}+\frac{1}{2}c_7	&	0		&	0 \\
		0		& 0	&	-\sqrt{c_7}\cos(\frac{\pi}{180}c_{8})	&	-\sqrt{c_7}\sin (\frac{\pi}{180}c_{8})	 \\
		0		& 0	&	\sqrt{c_7}\sin(\frac{\pi}{180}c_{8})	&	-\sqrt{c_7}\cos (\frac{\pi}{180}c_{8})	 \\
	\end{array} \right)}
$$
$$
	M_{s} = {\footnotesize\left( \begin{array}{c c c c}
		\frac{1}{2}+\frac{1}{2}c_9	&		\frac{1}{2}-\frac{1}{2}c_9	&	0		&	0 \\
		\frac{1}{2}-\frac{1}{2}c_9	&		\frac{1}{2}+\frac{1}{2}c_9	&	0		&	0 \\
		0		& 0	&	-\sqrt{c_9}\cos(\frac{\pi}{180}c_{10})	&	-\sqrt{c_9}\sin(\frac{\pi}{180}c_{10}) \\
		0		& 0	&	\sqrt{c_9}\sin(\frac{\pi}{180}c_{10})	&	-\sqrt{c_9}\cos(\frac{\pi}{180}c_{10})	 \\
	\end{array} \right)}
$$

where,
\begin{itemize}
	\item $c_0$, $c_1$, $c_2$, $c_3$ and $c_4$ are the parameters of the entrance lens.
	\item $c_5$ and $c_6$ are the de-attenuation and phase difference of the azimuth mirror.
	\item $c_7$ and $c_8$ are the de-attenuation and phase difference of the elevation mirror.
	\item $c_9$ and $c_{10}$ are the de-attenuation and phase difference of the Schupmann mirror.
	\item $c_{11}$ is the angle of the field mirror.
\end{itemize}
 We used a Levenberg-Marquardt algorithm \citep[see][]{2002nr} to fit the 12 parameters $(c)$ of the model to the calibration data. The orthogonality of 1-m polarizer states is maximum every $45^\circ$. Thus only four  angles are used in our fitting routine: $0^\circ,45^\circ,90^\circ, 135^\circ$. The quality of the fit does not improve substantially by including data from other angles. The derived parameters of the Mueller matrix of the telescope are shown in Table \ref{fitval}. Unfortunately, we have not been able to analyze the errors due to time constraints.
 \begin{table}
 \centering
 \caption{Parameters of the telescope model obtained from calibration data.}
 \begin{tabular}{c | l}
 	Parameter & Fitted value \\
 	\hline 
	$c_0$  &		$+9.596161\cdot 10^{-1}$		\\
	$c_1$  &		$-2.541791\cdot 10^{-3}$			\\
	$c_2$  &		$-8.337375\cdot 10^{-3}$	\\
	$c_3$  &		$+9.640901\cdot 10^{-1}$		\\
	$c_4$  &		$-1.875988\cdot 10^{-2}$		\\
	$c_5$  &		$+9.164327 \cdot 10^{-1}$		\\
	$c_6$  & 		 $+1.465364\cdot 10^1$\\
	$c_7$  &		$+1.017471\cdot 10^0$		\\
	$c_8$  &		$+1.326977\cdot 10^0$	\\
	$c_9$  &		 $+1.010035\cdot 10^0$	\\
	$c_{10}$ &		 $-1.592178\cdot 10^0$	\\
	$c_{11}$ &		 $+4.000000\cdot 10^0$	\\
\end{tabular}
	\label{fitval}
\end{table}
%However, it is important to point that the quality of the calibration data is determined by the quality of the 1 meter linear polarizer. In our case, the polarizer has a significant leak of unpolarized light at 854.2 nm given that the extinction ratio is \textbf{$0.4$}. 

The quality of the calibration data is limited by the quality of the 1-m linear polarizer. The problem is posed in such a way, that we cannot measure the extinction ratio of the polarizer and the parameters of the lens at the same time. In our case, the polarizer has a significant leak of unpolarized light at 854.2 nm. Using small samples of the sheets used to construct the 1-m polarizer, we have estimated the extinction ratio of the 1-m polarizer to be approximately $0.4$, so the parameters of the lens have been determined assuming that value. 

\begin{figure}[]
      \centering
      \resizebox{\hsize}{!}{\includegraphics[trim=0.2cm 0 0.2cm 0, clip]{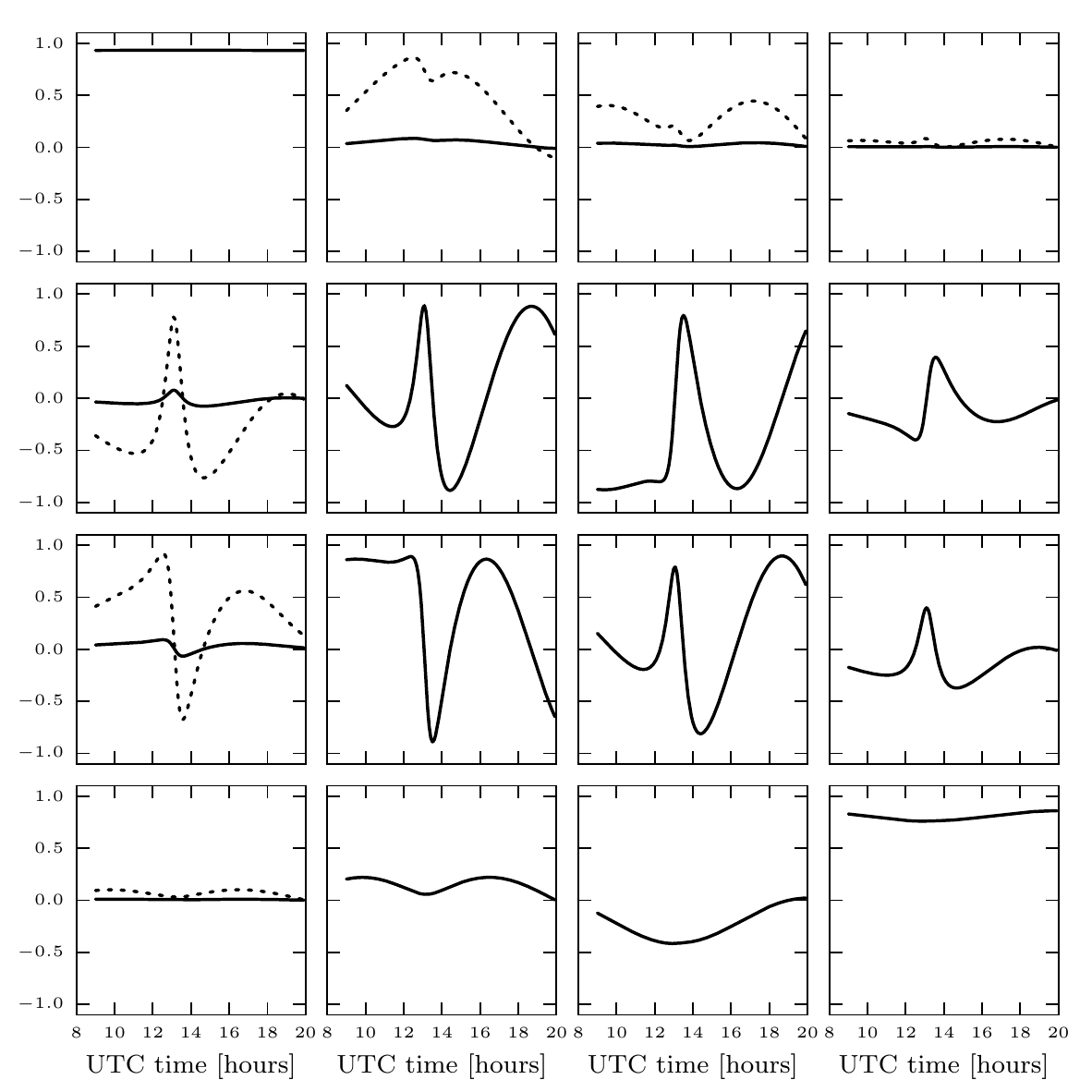}}
        \caption{
		Mueller matrix of the telescope computed for 2009-05-28. Each panel corresponds to one element of the matrix placed in the same order. The solid line represents the time dependence of each value. The dashed line is the same value multiplied by a factor 10.
       }
        \label{telmull}
\end{figure}

In Fig. \ref{telmull} the time dependence of the Mueller matrix of the telescope is shown along a whole day. The largest changes occur when the sun is close to zenith, because of the rapid movement of the telescope.
%\subsection{A reference for linear polarization}
 The linear polarization reference is defined by the first mirror after the lens, however this is not very useful in practice because the turret introduces image rotation along the day. Instead, we use the solar north as a reference for positive $Q$ by applying an extra rotation to $M_{tel}$. The rotation is produced by reflections inside the turret and by the variation of the angle between the first mirror after the entrance lens and the solar north along the day. The angle between the first mirror and the solar north $(\beta)$ is computed by the telescope software every 30 seconds. Eq. \ref{rotnor} transforms the reference of $Q$ and $U$ to solar North-South axis.
\begin{equation} 
 	M'_{tel} = M_{tel} \cdot R(\beta) \label{rotnor}
 \end{equation}
 
 The only remaining question is the location of the solar north in our science images. At this point, Stokes $Q$ and $U$ are relative to the solar North-South axis, but there is no coupling between the polarization calibration and the image orientation. The angle between solar north and the horizontal on the optical table is
 \begin{equation} 
	\omega = \varphi - \theta - \mbox{TC} - \beta,
	\label{nang}
 \end{equation}
 where $\varphi$ is the azimuth, $\theta$ is the elevation, TC is the \emph{table constant} and $\beta$ is the tilt angle between first mirror in the telescope and solar north. The table constant is relative to the orientation of the optical table and it is $+48^\circ$ for the current setup. 
 \begin{figure}[]
      \centering
      \resizebox{0.83\hsize}{!}{\includegraphics[]{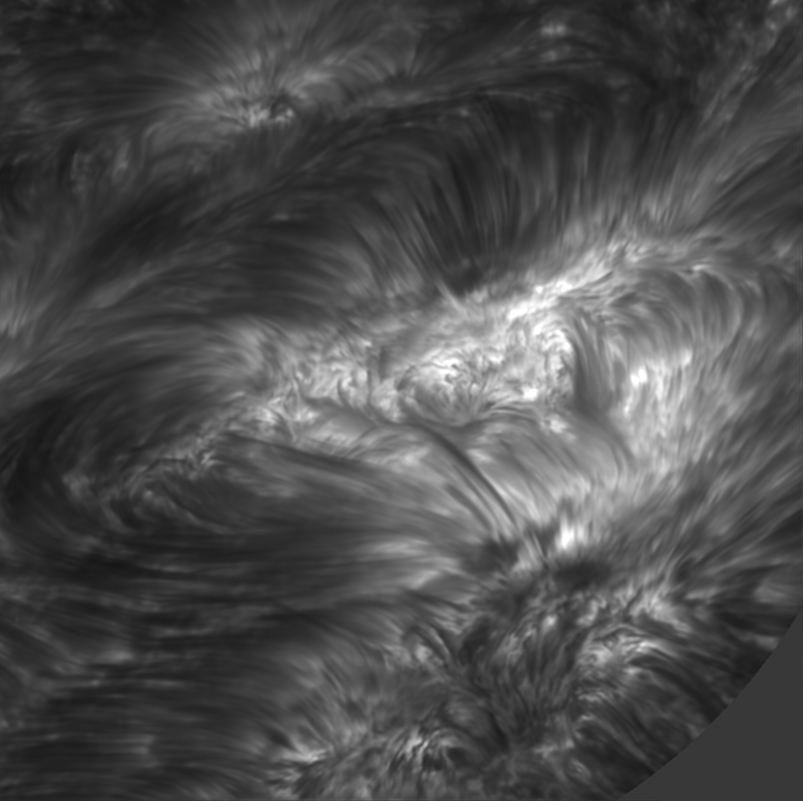}}
      \resizebox{0.83\hsize}{!}{\includegraphics[]{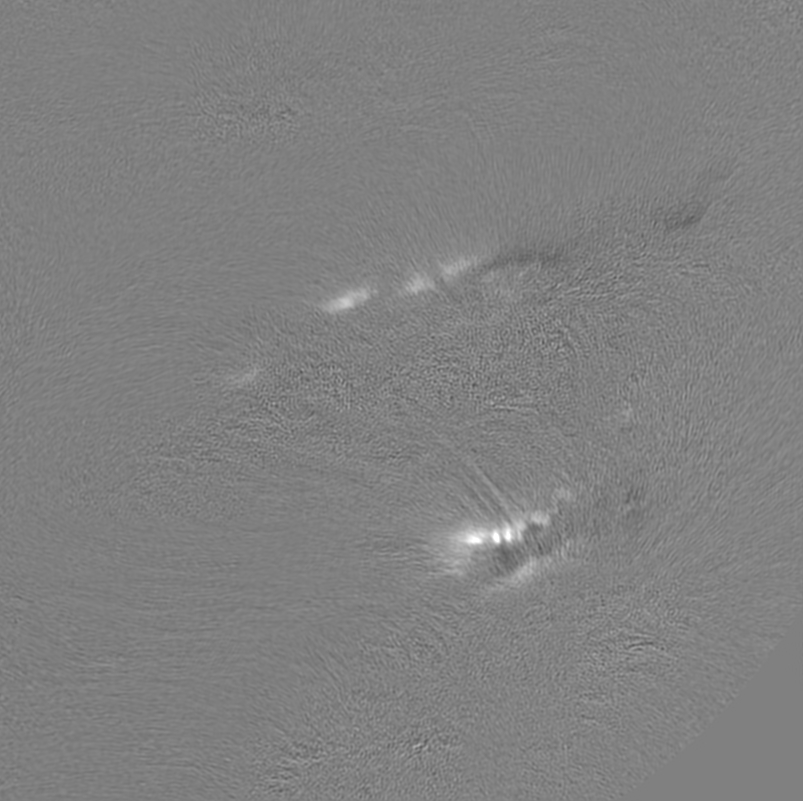}}
        \caption{
		Stokes $I$ (top) and $Q$ (bottom) monochromatic images acquired in \ion{Ca}{ii} 8542 \AA \ at $-161$ m\AA \ from the core of the line. Stokes $Q$ is scaled to $\pm 3\%$ of the continuum intensity.
       }
        \label{polres}
\end{figure}
 \begin{figure}[]
      \centering
      \resizebox{0.83\hsize}{!}{\includegraphics[]{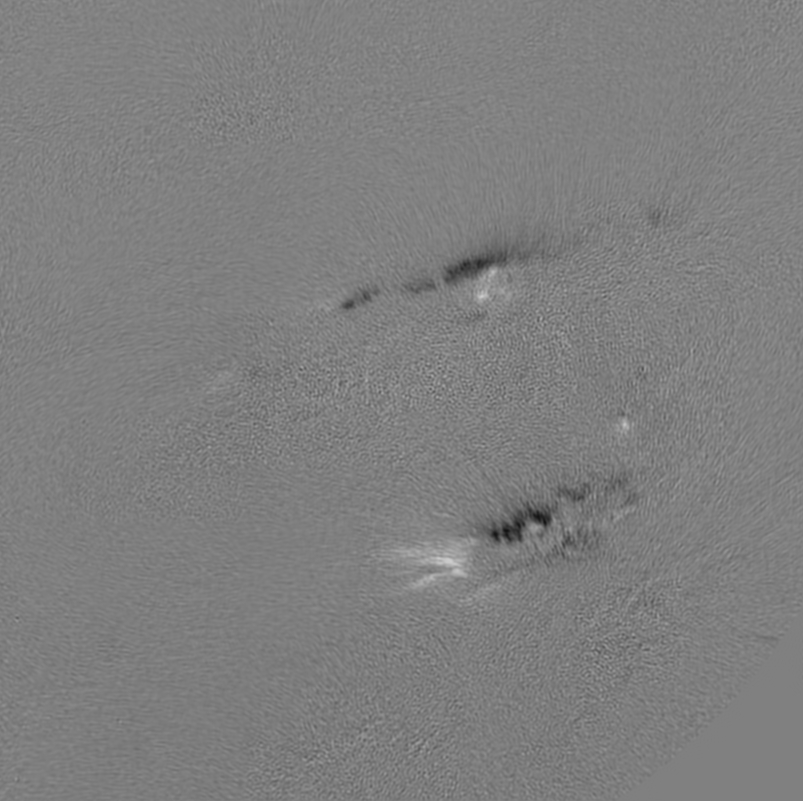}}
      \resizebox{0.83\hsize}{!}{\includegraphics[]{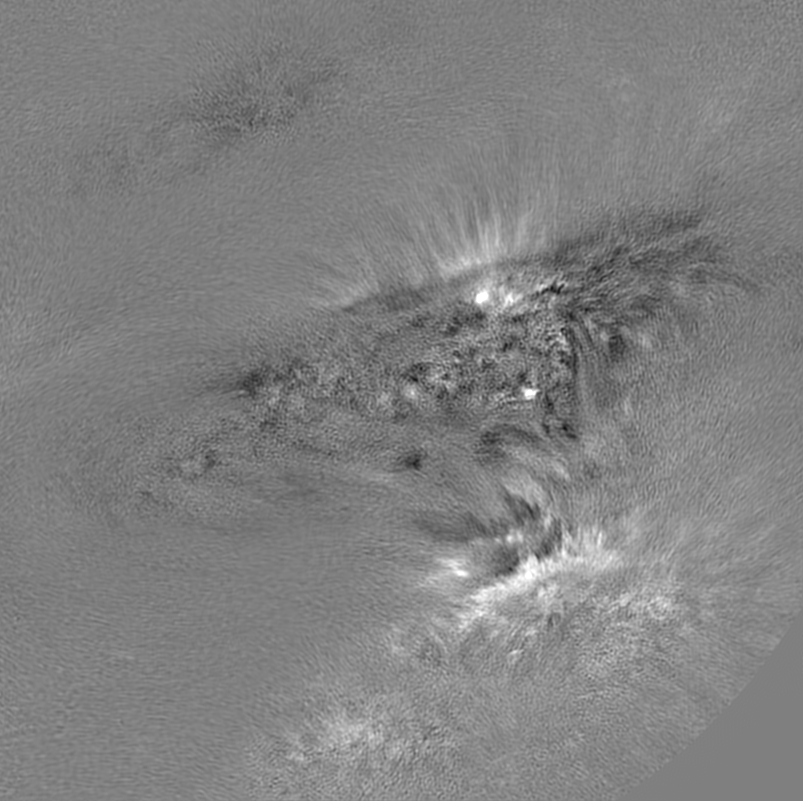}}
        \caption{
				Stokes $U$ (top) and $V$ (bottom) monochromatic images acquired in \ion{Ca}{ii} 8542 \AA \ at $-161$ m\AA \ from the core of the line. The images are scaled to $\pm 3\%$ of the continuum intensity.
       }
        \label{polres2}
\end{figure}

Fig. \ref{polres} and \ref{polres2} show monochromatic Stokes $I$, $Q$, $U$, $V$ images acquired in \ion{Ca}{ii} 8542 \AA. The dataset has been restored using the de-scattering scheme from \S\ref{backsc} and the telescope model presented in the current section. This dataset is used in \textbf{Paper III} to measure the alignment between fibrils and magnetic field.

\chapter{Summary of papers}
\section*{Paper I: Solar velocity references from 3D HD photospheric models}
The aim of this paper is to help observers to accurately calibrate line-of-sight velocities. We use a 3D hydrodynamic simulation of the solar photosphere to compute spatially-averaged spectra that can be used as absolute velocity references. The line profiles are computed at different heliocentric angles, from disk center towards the limb. 

Our synthetic profiles are compared with observational data and several experiments are computed to estimate the accuracy of our method, which has an estimated error of approximately $\pm50$ m s$^{-1}$ at disk center. Our tests suggest that the variation of the bisectors towards the limb, is mostly produced by the 3D topology of the photosphere.

In Paper I, I carried out all the calculations and prepared all figures. The collaborators contributed to the scientific discussion and assisted in the writing.

\section*{Paper II: Non-LTE inversions from a 3D MHD chromospheric model}
In Paper II, we create synthetic full-Stokes observations in \ion{Ca}{ii} 8542 \AA \ from a snapshot of a realistic 3D simulation of the solar atmosphere. These observations are used to estimate the amplitude of the Stokes profiles in quiet Sun. We discuss the effect that spectral degradation and noise have on our observations and discuss possible requirements for future instrumentation.

In the second part of the paper, we use our synthetic observations to test our non-LTE inversion code. The fitted model is compared with the quantities from the 3D snapshot. We conclude that the inversion code is able to estimate the average chromospheric value of magnetic field, line-of-sight and velocity. 3D non-LTE effects seem to affect the fitted temperature that in general presents less contrast than the original model.

My contribution to Paper II was to compute the full-Stokes simulated observations using the population densities provided by J. Leenaarts and carried out the Non-LTE inversions of the data. For that purpose I wrote an improved parallel version of NICOLE using MPI and a \emph{master-slave} scheme. I prepared the main structure of the text in the paper and created all the figures. 

\section*{Paper III: Are solar chromospheric fibrils tracing the magnetic field?}
The aim of this letter is to obtain an observational evidence that confirms the alignment between chromospheric fibrils and magnetic field. We use two datasets acquired with SST/CRISP and DST/SPINOR to measure the orientation of magnetic field along chromospheric fibrils. We find that many fibrils are aligned with magnetic field, however in both datasets there are evidences of misalignment in some cases. 

For this paper I provide a restored dataset from CRISP, compensated for telescope polarization and with a calibrated reference for Stokes $Q$ and $U$. My co-author prepared the SPINOR data and contributed to the scientific discussion and the writing.

\section*{Paper IV: Stokes imaging polarimetry using image restoration at the Swedish 1-m Solar Telescope II: A calibration strategy for Fabry-P\'erot based instruments}

The image restoration step that is applied to our data, decouples the 1-to-1 relation between the pixels of the CCD. When image reconstruction is applied to data showing sharp spatial variations produced by intrumentation, artifacts can appear. We propose a flat-fielding scheme for polarimetric data acquired with CRISP. We discuss the effect of the polarization introduced by the telescope and the optical setup in our flat-field data. In order to correct for spurious intensity fluctuations from cavity errors and reflectivity errors, we use a numerical framework that allows to model the spectral line on each pixel of the CCD.

My contribution to this paper was to implement the numerical scheme that is used to model the flat fields and remove the intensity fluctuations produced by cavity errors and reflectivity errors. I contributed to the scientific discussion and prepared some of the figures in the paper.

\chapter{Publications not included in this thesis}
\usecounter{papers}
\begin{itemize} 
\item \textbf{CRISP Spectropolarimetric Imaging of Penumbral Fine Structure}

Scharmer~G.~B., Narayan~G., Hillberg~T., de la Cruz Rodr\'iguez~J., L\"ofdahl~M.~G., Kiselman~D., S\"utterlin~P., van Noort M., Lagg~A., 2008, \apjl, 689, L69.
\\
\vspace{3mm}
\item \textbf{The magnetic SW Sextantis star RXJ1643.7+3402}

Rodr\'iguez-Gil P., Mart\'inez-Pais I. G., de la Cruz Rodr\'iguez J., 2009, \mnras, 395, 973.
\\
\vspace{3mm}
\item \textbf{High-order aberration compensation with Multi-frame Blind Deconvolution and Phase Diversity image restoration techniques}

Scharmer~G.~B., L\"ofdahl~M.~G., van Werkhoven T. I. M., de~la~Cruz~Rodr\'iguez~J., 2010, \aap, 521, A68
\\
\vspace{3mm}
\item \textbf{Observation and analysis of chromospheric magnetic fields}
	
de la Cruz Rodr\'iguez J., Socas-Navarro H., van Noort M., Rouppe van der Voort L., to appear in Proceedings of the 25th NSO Workshop: Chromospheric Structure and Dynamics, Memorie della Societa' Astronomica Italiana.
\\
  \end{itemize}
\vspace{13pt}

%\chapter{Summary of papers}
%\include{summary_of_papers}

\backmatter
\chapter{Acknowledgements}
I gratefully acknowledge the Institute for Solar Physics of the Royal Swedish Academy of Sciences and the USO-SP Graduate School for Solar Physics for giving me a position in science in a prosperous academic environment.  I  especially value the scientific discussions with my collaborators \textbf{Hector Socas-Navarro}, \textbf{Michiel van Noort} and \textbf{Roald Schnerr} from whom I learned so much.  I have benefited from the assistance and advice provided by my supervisors \textbf{Dan Kiselman, G\"oran Scharmer} and \textbf{Mats Carlsson}. I also appreciate the help from \textbf{Mats L\"ofdahl} who kindly commented on the manuscript of my thesis.

My thanks to \textbf{Tiago Pereira} and to the \textbf{Institute of theoretical Astrophysics} of Oslo for providing  the 3D simulations used in my research. I especially enjoyed being in contact with  \textbf{Luc~Rouppe~van~der~Voort} who shared observational data and also his knowledge with me. 
I also acknowledge \textbf{Pit~S\"utterlin} and \textbf{Rolf~Kever} who have been of great help at the SST on La Palma. Here in Stockholm it has been great to share the office (and pubs) with \textbf{Vasco~Henriques}.

I gratefully acknowledge the financial support of the European Commission during the first three years of my PhD. Many thanks to my institute for extending my financial support during the last months of my research.

\begin{figure}[]
    	\centering
        \resizebox{0.3\hsize}{!}{\includegraphics[]{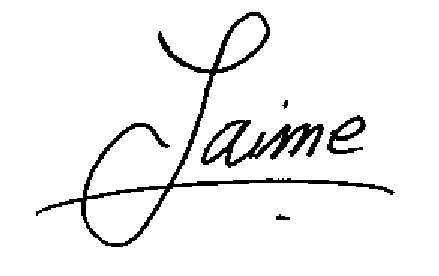}}
\end{figure}

%\chapter{Papers and proceedings not included in this thesis}

%\nocite{*}
\bibliographystyle{aa}
\bibliography{references}

\begin{thebibliography}{77}
\expandafter\ifx\csname natexlab\endcsname\relax\def\natexlab#1{#1}\fi

\bibitem[{{Asensio Ramos} \& {Trujillo Bueno}(2009)}]{2009asensio-ramos}
{Asensio Ramos}, A. \& {Trujillo Bueno}, J. 2009, in Astronomical Society of
  the Pacific Conference Series, Vol. 405, Solar Polarization 5: In Honor of
  Jan Stenflo, ed. {S.~V.~Berdyugina and K.~N.~Nagendra and R.~Ramelli}, 281

\bibitem[{{Asplund} {et~al.}(2000){Asplund}, {Nordlund}, {Trampedach}, \&
  {Stein}}]{2000asplund2}
{Asplund}, M., {Nordlund}, {\AA}., {Trampedach}, R., \& {Stein}, R.~F. 2000,
  \aap, 359, 743

\bibitem[{{Bahng} \& {Schwarzschild}(1961)}]{1961bahng}
{Bahng}, J. \& {Schwarzschild}, M. 1961, \apj, 134, 312

\bibitem[{{Balthasar}(1988)}]{1988balthasar}
{Balthasar}, H. 1988, \aaps, 72, 473

\bibitem[{{Beckers}(1977)}]{1977beckers}
{Beckers}, J.~M. 1977, \apj, 213, 900

\bibitem[{{Bellot Rubio} {et~al.}(2004){Bellot Rubio}, {Balthasar}, \&
  {Collados}}]{2004bellot-rubio}
{Bellot Rubio}, L.~R., {Balthasar}, H., \& {Collados}, M. 2004, \aap, 427, 319

\bibitem[{{Bellot Rubio} \& {Borrero}(2002)}]{2002bellot-rubio2}
{Bellot Rubio}, L.~R. \& {Borrero}, J.~M. 2002, \aap, 391, 331

\bibitem[{{Bellot Rubio} {et~al.}(2008){Bellot Rubio}, {Tritschler}, \&
  {Mart{\'{\i}}nez Pillet}}]{2008bellot-rubio}
{Bellot Rubio}, L.~R., {Tritschler}, A., \& {Mart{\'{\i}}nez Pillet}, V. 2008,
  \apj, 676, 698

\bibitem[{{Borrero} \& {Bellot Rubio}(2002)}]{2002bellot-rubio}
{Borrero}, J.~M. \& {Bellot Rubio}, L.~R. 2002, \aap, 385, 1056

\bibitem[{{Brault} \& {Neckel}(1987)}]{fts-atlas}
{Brault}, J.~W. \& {Neckel}, H. 1987, Spectral Atlas of Solar Absolute
  Disk-averaged and Disk-Center Intensity from 3290 to 12510 \AA,
  ftp://ftp.hs.uni-hamburg.de/pub/outgolng/FTS-Atlas

\bibitem[{{Cacciani} {et~al.}(2006){Cacciani}, {Briguglio}, {Massa}, \&
  {Rapex}}]{2006cacciani}
{Cacciani}, A., {Briguglio}, R., {Massa}, F., \& {Rapex}, P. 2006, Celestial
  Mechanics and Dynamical Astronomy, 95, 425

\bibitem[{{Carlsson} {et~al.}(2010){Carlsson}, {Hansteen}, \&
  {Gudiksen}}]{2010carlsson}
{Carlsson}, M., {Hansteen}, V.~H., \& {Gudiksen}, B.~V. 2010, ArXiv e-prints
  1001.1546

\bibitem[{{Carlsson} \& {Stein}(1995)}]{1995carlsson}
{Carlsson}, M. \& {Stein}, R.~F. 1995, \apjl, 440, L29

\bibitem[{{Cauzzi} {et~al.}(2009){Cauzzi}, {Reardon}, {Rutten}, {Tritschler},
  \& {Uitenbroek}}]{2009cauzzi}
{Cauzzi}, G., {Reardon}, K., {Rutten}, R.~J., {Tritschler}, A., \&
  {Uitenbroek}, H. 2009, \aap, 503, 577

\bibitem[{{Cauzzi} {et~al.}(2008){Cauzzi}, {Reardon}, {Uitenbroek},
  {Cavallini}, {Falchi}, {Falciani}, {Janssen}, {Rimmele}, {Vecchio}, \&
  {W{\"o}ger}}]{2008cauzzi}
{Cauzzi}, G., {Reardon}, K.~P., {Uitenbroek}, H., {et~al.} 2008, \aap, 480, 515

\bibitem[{{Cauzzi} {et~al.}(2007){Cauzzi}, {Reardon}, {Vecchio}, {Janssen}, \&
  {Rimmele}}]{2007cauzzi}
{Cauzzi}, G., {Reardon}, K.~P., {Vecchio}, A., {Janssen}, K., \& {Rimmele}, T.
  2007, in Astronomical Society of the Pacific Conference Series, Vol. 368, The
  Physics of Chromospheric Plasmas, ed. {P.~Heinzel, I.~Dorotovi{\v c}, \&
  R.~J.~Rutten}, 127

\bibitem[{{Centeno} {et~al.}(2008){Centeno}, {Trujillo Bueno}, {Uitenbroek}, \&
  {Collados}}]{2008centeno}
{Centeno}, R., {Trujillo Bueno}, J., {Uitenbroek}, H., \& {Collados}, M. 2008,
  \apj, 677, 742

\bibitem[{{Cheung} {et~al.}(2007){Cheung}, {Sch{\"u}ssler}, \&
  {Moreno-Insertis}}]{2007cheung}
{Cheung}, M.~C.~M., {Sch{\"u}ssler}, M., \& {Moreno-Insertis}, F. 2007, \aap,
  461, 1163

\bibitem[{{de la Cruz Rodr{\'{\i}}guez} {et~al.}(2010){de la Cruz
  Rodr{\'{\i}}guez}, {Socas-Navarro}, {van Noort}, \& {Rouppe van der
  Voort}}]{2010delacruz}
{de la Cruz Rodr{\'{\i}}guez}, J., {Socas-Navarro}, H., {van Noort}, M., \&
  {Rouppe van der Voort}, L. 2010, ArXiv e-prints 1004.0698

\bibitem[{{de la Cruz Rodr\'iguez} {et~al.}(2011){de la Cruz Rodr\'iguez}, {van
  Noort}, \& {Schnerr}}]{2011delacruz}
{de la Cruz Rodr\'iguez}, J., {van Noort}, M., \& {Schnerr}, R. 2011, In
  preparation

\bibitem[{{De Pontieu} {et~al.}(2007){De Pontieu}, {McIntosh}, {Hansteen},
  {Carlsson}, {Schrijver}, {Tarbell}, {Title}, {Shine}, {Suematsu}, {Tsuneta},
  {Katsukawa}, {Ichimoto}, {Shimizu}, \& {Nagata}}]{2007depontieu}
{De Pontieu}, B., {McIntosh}, S., {Hansteen}, V.~H., {et~al.} 2007, \pasj, 59,
  655

\bibitem[{{Dravins} {et~al.}(1981){Dravins}, {Lindegren}, \&
  {Nordlund}}]{1981dravins}
{Dravins}, D., {Lindegren}, L., \& {Nordlund}, A. 1981, \aap, 96, 345

\bibitem[{{Fossum} \& {Carlsson}(2005)}]{2005fossum}
{Fossum}, A. \& {Carlsson}, M. 2005, in ESA Special Publication, Vol. 600, The
  Dynamic Sun: Challenges for Theory and Observations

\bibitem[{{Franz} \& {Schlichenmaier}(2009)}]{2009franz}
{Franz}, M. \& {Schlichenmaier}, R. 2009, \aap, 508, 1453

\bibitem[{{Hansteen} {et~al.}(2007){Hansteen}, {Carlsson}, \&
  {Gudiksen}}]{2007hansteen}
{Hansteen}, V.~H., {Carlsson}, M., \& {Gudiksen}, B. 2007, in Astronomical
  Society of the Pacific Conference Series, Vol. 368, The Physics of
  Chromospheric Plasmas, ed. {P.~Heinzel, I.~Dorotovi{\v c}, \& R.~J.~Rutten},
  107

\bibitem[{{Hansteen} {et~al.}(2006){Hansteen}, {De Pontieu}, {Rouppe van der
  Voort}, {van Noort}, \& {Carlsson}}]{2006hansteen}
{Hansteen}, V.~H., {De Pontieu}, B., {Rouppe van der Voort}, L., {van Noort},
  M., \& {Carlsson}, M. 2006, \apjl, 647, L73

\bibitem[{{Judge}(2006)}]{2006judge}
{Judge}, P. 2006, in Astronomical Society of the Pacific Conference Series,
  Vol. 354, Solar MHD Theory and Observations: A High Spatial Resolution
  Perspective, ed. {J.~Leibacher, R.~F.~Stein, \& H.~Uitenbroek}, 259

\bibitem[{{Judge} {et~al.}(2010){Judge}, {Tritschler}, {Uitenbroek}, {Reardon},
  {Cauzzi}, \& {de Wijn}}]{2010judge}
{Judge}, P.~G., {Tritschler}, A., {Uitenbroek}, H., {et~al.} 2010, \apj, 710,
  1486

\bibitem[{{Kahn}(1961)}]{1961kahn}
{Kahn}, F.~D. 1961, \apj, 134, 343

\bibitem[{{Kuckein} {et~al.}(2010){Kuckein}, {Centeno}, \& {Martinez
  Pillet}}]{2010kuckein}
{Kuckein}, C., {Centeno}, R., \& {Martinez Pillet}, V. 2010, ArXiv e-prints
  1010.4260

\bibitem[{{Lagg} {et~al.}(2004){Lagg}, {Woch}, {Krupp}, \&
  {Solanki}}]{2004lagg}
{Lagg}, A., {Woch}, J., {Krupp}, N., \& {Solanki}, S.~K. 2004, \aap, 414, 1109

\bibitem[{{Langangen} {et~al.}(2008){Langangen}, {Carlsson}, {Rouppe van der
  Voort}, {Hansteen}, \& {De Pontieu}}]{2008langangen}
{Langangen}, {\O}., {Carlsson}, M., {Rouppe van der Voort}, L., {Hansteen}, V.,
  \& {De Pontieu}, B. 2008, \apj, 673, 1194

\bibitem[{{Langangen} {et~al.}(2007){Langangen}, {Carlsson}, {Rouppe van der
  Voort}, \& {Stein}}]{2007langangen}
{Langangen}, {\O}., {Carlsson}, M., {Rouppe van der Voort}, L., \& {Stein},
  R.~F. 2007, \apj, 655, 615

\bibitem[{{Leenaarts} {et~al.}(2009){Leenaarts}, {Carlsson}, {Hansteen}, \&
  {Rouppe van der Voort}}]{2009leenaarts}
{Leenaarts}, J., {Carlsson}, M., {Hansteen}, V., \& {Rouppe van der Voort}, L.
  2009, \apjl, 694, L128

\bibitem[{{Leenaarts} {et~al.}(2007){Leenaarts}, {Carlsson}, {Hansteen}, \&
  {Rutten}}]{2007leenaarts}
{Leenaarts}, J., {Carlsson}, M., {Hansteen}, V., \& {Rutten}, R.~J. 2007, \aap,
  473, 625

\bibitem[{{Leighton} {et~al.}(1962){Leighton}, {Noyes}, \& {Simon}}]{1962dop}
{Leighton}, R.~B., {Noyes}, R.~W., \& {Simon}, G.~W. 1962, \apj, 135, 474

\bibitem[{{Manso Sainz} \& {Trujillo Bueno}(2010)}]{2010manso}
{Manso Sainz}, R. \& {Trujillo Bueno}, J. 2010, \apj, 722, 1416

\bibitem[{{Martinez Pillet} {et~al.}(1997){Martinez Pillet}, {Lites}, \&
  {Skumanich}}]{1997pillet}
{Martinez Pillet}, V., {Lites}, B.~W., \& {Skumanich}, A. 1997, \apj, 474, 810

\bibitem[{{Meyer} \& {Schmidt}(1968)}]{1968meyer}
{Meyer}, F. \& {Schmidt}, H.~U. 1968, Mitteilungen der Astronomischen
  Gesellschaft Hamburg, 25, 194

\bibitem[{{Montesinos} \& {Thomas}(1997)}]{1997montesinos}
{Montesinos}, B. \& {Thomas}, J.~H. 1997, \nat, 390, 485

\bibitem[{{Ortiz} {et~al.}(2010){Ortiz}, {Bellot Rubio}, \& {Rouppe van der
  Voort}}]{2010ortiz}
{Ortiz}, A., {Bellot Rubio}, L.~R., \& {Rouppe van der Voort}, L. 2010, \apj,
  713, 1282

\bibitem[{{Pietarila} {et~al.}(2007{\natexlab{a}}){Pietarila}, {Socas-Navarro},
  \& {Bogdan}}]{2007pietarila}
{Pietarila}, A., {Socas-Navarro}, H., \& {Bogdan}, T. 2007{\natexlab{a}}, \apj,
  670, 885

\bibitem[{{Pietarila} {et~al.}(2007{\natexlab{b}}){Pietarila}, {Socas-Navarro},
  \& {Bogdan}}]{2007pietarila2}
{Pietarila}, A., {Socas-Navarro}, H., \& {Bogdan}, T. 2007{\natexlab{b}}, in
  Astronomical Society of the Pacific Conference Series, Vol. 368, The Physics
  of Chromospheric Plasmas, ed. {P.~Heinzel, I.~Dorotovi{\v c}, \&
  R.~J.~Rutten}, 139

\bibitem[{Press {et~al.}(2002)Press, Teukolsky, Vetterling, \&
  Flannery}]{2002nr}
Press, W.~H., Teukolsky, S.~A., Vetterling, W.~T., \& Flannery, B.~P. 2002,
  {Numerical recipes in C++: The art of scientific computing}, 2nd edn.
  (Cambridge University Press)

\bibitem[{{Priest}(1982)}]{1982priest}
{Priest}, E.~R. 1982, Geophysics and Astrophysics Monographs, Vol.~21, {Solar
  magneto-hydrodynamics} (Dordrecht: D. Reidel Pub.~Co.)

\bibitem[{{Rouppe van der Voort} {et~al.}(2009){Rouppe van der Voort},
  {Leenaarts}, {De Pontieu}, {Carlsson}, \& {Vissers}}]{2009rouppe}
{Rouppe van der Voort}, L., {Leenaarts}, J., {De Pontieu}, B., {Carlsson}, M.,
  \& {Vissers}, G. 2009, \apj, 705, 272

\bibitem[{{Rouppe van der Voort} {et~al.}(2007){Rouppe van der Voort}, {De
  Pontieu}, {Hansteen}, {Carlsson}, \& {van Noort}}]{2007rouppe}
{Rouppe van der Voort}, L.~H.~M., {De Pontieu}, B., {Hansteen}, V.~H.,
  {Carlsson}, M., \& {van Noort}, M. 2007, \apjl, 660, L169

\bibitem[{{Ruiz Cobo} \& {del Toro Iniesta}(1992)}]{1992ruiz-cobo}
{Ruiz Cobo}, B. \& {del Toro Iniesta}, J.~C. 1992, \apj, 398, 375

\bibitem[{{Rutten}(2006)}]{2006rutten}
{Rutten}, R.~J. 2006, in Astronomical Society of the Pacific Conference Series,
  Vol. 354, Solar MHD Theory and Observations: A High Spatial Resolution
  Perspective, ed. {J.~Leibacher, R.~F.~Stein, \& H.~Uitenbroek}, 276

\bibitem[{{Rutten}(2007)}]{2007rutten}
{Rutten}, R.~J. 2007, in Astronomical Society of the Pacific Conference Series,
  Vol. 368, The Physics of Chromospheric Plasmas, ed. {P.~Heinzel,
  I.~Dorotovi{\v c}, \& R.~J.~Rutten}, 27

\bibitem[{{Sanchez Almeida} \& {Lites}(1992)}]{1992sanchez-almeida}
{Sanchez Almeida}, J. \& {Lites}, B.~W. 1992, \apj, 398, 359

\bibitem[{{Scharmer}(2006)}]{2006scharmer}
{Scharmer}, G.~B. 2006, \aap, 447, 1111

\bibitem[{{Scharmer}(2008)}]{2008scharmerR}
{Scharmer}, G.~B. 2008, Physica Scripta Volume T, 133, 014015

\bibitem[{{Scharmer} {et~al.}(2003){Scharmer}, {Bjelksj\"o}, {Korhonen},
  {Lindberg}, \& {Petterson}}]{2003scharmer}
{Scharmer}, G.~B., {Bjelksj\"o}, K., {Korhonen}, T.~K., {Lindberg}, B., \&
  {Petterson}, B. 2003, in Society of Photo-Optical Instrumentation Engineers
  (SPIE) Conference Series, Vol. 4853, Innovative Telescopes and
  Instrumentation for Solar Astrophysics, ed. {S.~L.~Keil \& S.~V.~Avakyan},
  341--350

\bibitem[{{Scharmer} {et~al.}(2002){Scharmer}, {Gudiksen}, {Kiselman},
  {L{\"o}fdahl}, \& {Rouppe van der Voort}}]{2002scharmer}
{Scharmer}, G.~B., {Gudiksen}, B.~V., {Kiselman}, D., {L{\"o}fdahl}, M.~G., \&
  {Rouppe van der Voort}, L.~H.~M. 2002, \nat, 420, 151

\bibitem[{{Scharmer} {et~al.}(2008){Scharmer}, {Narayan}, {Hillberg}, {de la
  Cruz Rodriguez}, {L{\"o}fdahl}, {Kiselman}, {S{\"u}tterlin}, {van Noort}, \&
  {Lagg}}]{2008scharmer}
{Scharmer}, G.~B., {Narayan}, G., {Hillberg}, T., {et~al.} 2008, \apjl, 689,
  L69

\bibitem[{{Scharmer} \& {Spruit}(2006)}]{2006scharmerspruit}
{Scharmer}, G.~B. \& {Spruit}, H.~C. 2006, \aap, 460, 605

\bibitem[{{Selbing}(2005)}]{2005selbing}
{Selbing}, J. 2005, Master's thesis, Stockholm University

\bibitem[{{Shchukina} \& {Trujillo Bueno}(2001)}]{2001trujillo}
{Shchukina}, N. \& {Trujillo Bueno}, J. 2001, \apj, 550, 970

\bibitem[{{Socas-Navarro} {et~al.}(2010){Socas-Navarro}, {de la Cruz
  Rodr\'iguez}, {Asensio-Ramos}, {Trujillo-Bueno}, \& {Ruiz-Cobo}}]{nicoleref}
{Socas-Navarro}, H., {de la Cruz Rodr\'iguez}, J., {Asensio-Ramos}, A.,
  {Trujillo-Bueno}, J., \& {Ruiz-Cobo}, B. 2010, in preperation

\bibitem[{{Socas-Navarro} {et~al.}(2006){Socas-Navarro}, {Elmore}, {Pietarila},
  {Darnell}, {Lites}, {Tomczyk}, \& {Hegwer}}]{2006socas-navarro}
{Socas-Navarro}, H., {Elmore}, D., {Pietarila}, A., {et~al.} 2006, \solphys,
  235, 55

\bibitem[{{Socas-Navarro} {et~al.}(2000{\natexlab{a}}){Socas-Navarro},
  {Trujillo Bueno}, \& {Ruiz Cobo}}]{2000socas-navarro}
{Socas-Navarro}, H., {Trujillo Bueno}, J., \& {Ruiz Cobo}, B.
  2000{\natexlab{a}}, Science, 288, 1396

\bibitem[{{Socas-Navarro} {et~al.}(2000{\natexlab{b}}){Socas-Navarro},
  {Trujillo Bueno}, \& {Ruiz Cobo}}]{2000socas-navarro0}
{Socas-Navarro}, H., {Trujillo Bueno}, J., \& {Ruiz Cobo}, B.
  2000{\natexlab{b}}, \apj, 530, 977

\bibitem[{{Solanki} \& {Montavon}(1993)}]{1993solanki}
{Solanki}, S.~K. \& {Montavon}, C.~A.~P. 1993, \aap, 275, 283

\bibitem[{{Stein} \& {Nordlund}(1998)}]{1998stein}
{Stein}, R.~F. \& {Nordlund}, A. 1998, \apj, 499, 914

\bibitem[{{Stix}(2002)}]{2002stix}
{Stix}, M. 2002, {The sun: an introduction}, ed. {Stix, M.}

\bibitem[{{Thomas} {et~al.}(2002){Thomas}, {Weiss}, {Tobias}, \&
  {Brummell}}]{2002thomas}
{Thomas}, J.~H., {Weiss}, N.~O., {Tobias}, S.~M., \& {Brummell}, N.~H. 2002,
  \nat, 420, 390

\bibitem[{{Tritschler} {et~al.}(2004){Tritschler}, {Schlichenmaier}, {Bellot
  Rubio}, {the KAOS Team}, {Berkefeld}, \& {Schelenz}}]{2004tritschler}
{Tritschler}, A., {Schlichenmaier}, R., {Bellot Rubio}, L.~R., {et~al.} 2004,
  \aap, 415, 717

\bibitem[{{van Noort} {et~al.}(2005){van Noort}, {Rouppe van der Voort}, \&
  {L{\"o}fdahl}}]{2005noort}
{van Noort}, M., {Rouppe van der Voort}, L., \& {L{\"o}fdahl}, M.~G. 2005,
  \solphys, 228, 191

\bibitem[{{van Noort} \& {Rouppe van der Voort}(2006)}]{2006noort2}
{van Noort}, M.~J. \& {Rouppe van der Voort}, L.~H.~M. 2006, \apjl, 648, L67

\bibitem[{{Vecchio} {et~al.}(2009){Vecchio}, {Cauzzi}, \&
  {Reardon}}]{2009vecchio}
{Vecchio}, A., {Cauzzi}, G., \& {Reardon}, K.~P. 2009, \aap, 494, 269

\bibitem[{{Vernazza} {et~al.}(1981){Vernazza}, {Avrett}, \&
  {Loeser}}]{1981vernazza}
{Vernazza}, J.~E., {Avrett}, E.~H., \& {Loeser}, R. 1981, \apjs, 45, 635

\bibitem[{{Wedemeyer-B{\"o}hm} {et~al.}(2009){Wedemeyer-B{\"o}hm}, {Lagg}, \&
  {Nordlund}}]{2009wedemeyer}
{Wedemeyer-B{\"o}hm}, S., {Lagg}, A., \& {Nordlund}, {\AA}. 2009, \ssr, 144,
  317

\bibitem[{{Wedemeyer-B{\"o}hm} {et~al.}(2007){Wedemeyer-B{\"o}hm}, {Steiner},
  {Bruls}, \& {Rammacher}}]{2007wedemeyer}
{Wedemeyer-B{\"o}hm}, S., {Steiner}, O., {Bruls}, J., \& {Rammacher}, W. 2007,
  in Astronomical Society of the Pacific Conference Series, Vol. 368, The
  Physics of Chromospheric Plasmas, ed. {P.~Heinzel, I.~Dorotovi{\v c}, \&
  R.~J.~Rutten}, 93

\bibitem[{{Weiss} {et~al.}(2004){Weiss}, {Thomas}, {Brummell}, \&
  {Tobias}}]{2004weiss}
{Weiss}, N.~O., {Thomas}, J.~H., {Brummell}, N.~H., \& {Tobias}, S.~M. 2004,
  \apj, 600, 1073

\bibitem[{{Westendorp Plaza} {et~al.}(1997){Westendorp Plaza}, {del Toro
  Iniesta}, {Ruiz Cobo}, {Martinez Pillet}, {Lites}, \&
  {Skumanich}}]{1997westendorp-plaza}
{Westendorp Plaza}, C., {del Toro Iniesta}, J.~C., {Ruiz Cobo}, B., {et~al.}
  1997, \nat, 389, 47

\bibitem[{{Wilson} \& {Maskelyne}(1774)}]{1774wilson}
{Wilson}, A. \& {Maskelyne}, N. 1774, Royal Society of London Philosophical
  Transactions Series I, 64, 1

\end{thebibliography}

%\chapter{Paper I}
%\includepdf[pagecommand={},pages=-]{papers/2010bisec_cor.pdf}

\end{document}